\pdfoutput=1 
\documentclass{JINST}

\title{Study of Light Detection and Sensitivity for a Ton-scale Liquid Xenon Dark Matter Detector}

\author{Yuehuan Wei, Qing Lin, Xiang Xiao and Kaixuan Ni\\
\llap  INPAC, Department of Physics and Shanghai Key Laboratory for Particle Physics and Cosmology\\
Shanghai Jiao Tong University, 800 Dongchuan Road, Shanghai, 200240, P. R. China\\
  E-mail: \email{nikx@sjtu.edu.cn}}

\abstract{Ton-scale liquid xenon detectors operated in two-phase mode are
proposed and being constructed recently to explore the favored parameter space for
the Weakly Interacting Massive Particles (WIMPs) dark matter. To achieve
a better light collection efficiency while limiting the number of electronics
channels compared to the previous generation detectors, large-size photo-multiplier tubes (PMTs)
such as the 3-inch-diameter R11410 from Hamamatsu are suggested to replace
the 1-inch-square R8520 PMTs. In a two-phase xenon dark
matter detector, two PMT arrays on the top and bottom are usually used. In this study, we compare the performance of two different ton-scale liquid xenon detector configurations
with the same number of either R11410 (config.1) or R8520 (config.2) for the top PMT array,
while both using R11410 PMTs for the bottom array.
The self-shielding of liquid xenon suppresses the background from the PMTs and the dominant background is from the $pp$ solar neutrinos in the central fiducial volume.
The light collection efficiency for the primary scintillation light is largely affected by the xenon purity and the reflectivity of the reflectors.
In the optimistic situation with a 10 m light absorption length and a 95\% reflectivity,
the light collection efficiency is 43\%(34\%) for config.1(config.2).
In the conservative situation with a 2.5 m light absorption length and a 85\% reflectivity, the value is only 18\%(13\%) for config.1(config.2).
The difference between the two configurations is due to the larger PMT coverage on the top for config.1.
The slightly different position resolutions for the two configurations have a negligible effect on the sensitivity. Based on the above considerations,
we estimate the sensitivity reach of the two detector configurations.
Both configurations can reach a sensitivity of $2\sim3 \times 10^{-47} \rm cm^2$ for spin-independent WIMP-nucleon cross section for 100 GeV/$c^2$ WIMPs
after two live-years of operation.
The one with R8520 PMTs for the top PMT array is more cost-effective,
while the one with R11410 PMTs on the top has a factor of two better sensitivity for light WIMPs at 10 GeV/$c^2$.}


\begin{document}

\section{Introduction}
Direct detection of dark matter is of great interests in both particle physics
and astrophysics. Many experimental groups use different techniques to search
for WIMPs \cite{bib1, bib2}, postulated by theories beyond the standard model of particle physics.
Remarkable progress has been made in recent years, but the compatibility between
signal claims and null results of different experiments is far from being a settled
issue \cite{bib3, bib4, bib5, bib6, bib7}. The liquid xenon (LXe) technique is very competitive among various detection
techniques owning to the large mass, self-shielding ability, scalability, high scintillation
and ionization yields. Both primary scintillation light and secondary scintillation light from ionized
electrons drifting from the liquid to the gas phase, from a WIMP elastic scattering on the target nucleus,
are detected and used for energy and position reconstruction, as well as for particle identification \cite{bib6}.

To fully explore the favored parameter space for WIMPs dark matter in search of a robust and statistically significant signal,
ton-scale LXe detectors operated in two-phase mode were proposed \cite{bib8, bib9}.
For the time projection chamber (TPC) of a ton-scale detector, large PMT coverage will be used to optimize the light
detection which has a big impact on the energy threshold of the detector. To limit
the number of electronics channels for the PMT readout, large size R11410 PMTs~\cite{bib13,bib19} are
considered for the ton-scale detectors. However, the R11410 PMT is more radioactive
than the R8520 PMT used in the previous generation detectors such as XENON100 \cite{bib6}.
In this work, we compare two different detector configurations with the
same number of either R11410 (config.1) or R8520 (config.2) PMTs for the top array, while both using R11410 PMTs for the bottom PMT array.
The performance in terms of background contribution, light collection efficiency and position sensitivity were simulated in Geant4 \cite{bib10}.
In Sec.2, we describe these two different configurations.
The simulated background is discussed in Sec.3.
Light collection efficiency is presented in Sec.4.
The position reconstruction and its effect on the background are discussed in Sec.5.
Base on these considerations, we compare the projected sensitivity to dark matter for these two configurations in Sec.6.

\section{The conceptual design of the time projection chamber}
A schematic view of the two-phase xenon TPC is shown in figure 1.
A particle interaction in LXe produces scintillation light and ionization electrons.
The electrons drift towards the liquid-gas interface under the drift field between the grid and the cathode.
They are extracted into the gas xenon (GXe) under a strong electric field above the liquid which
is produced between the grid and anode at a few mm above the liquid-gas interface.
For a field larger than 10 kV/cm in the GXe, the extraction yield of electrons is close to 100\% \cite{bib11}.
Both the direct (S1) and the proportional light (S2) are detected by the PMT arrays installed at the top and bottom of the TPC.

\begin{figure}[htd]
\centerline{\includegraphics[width=.5\textwidth]{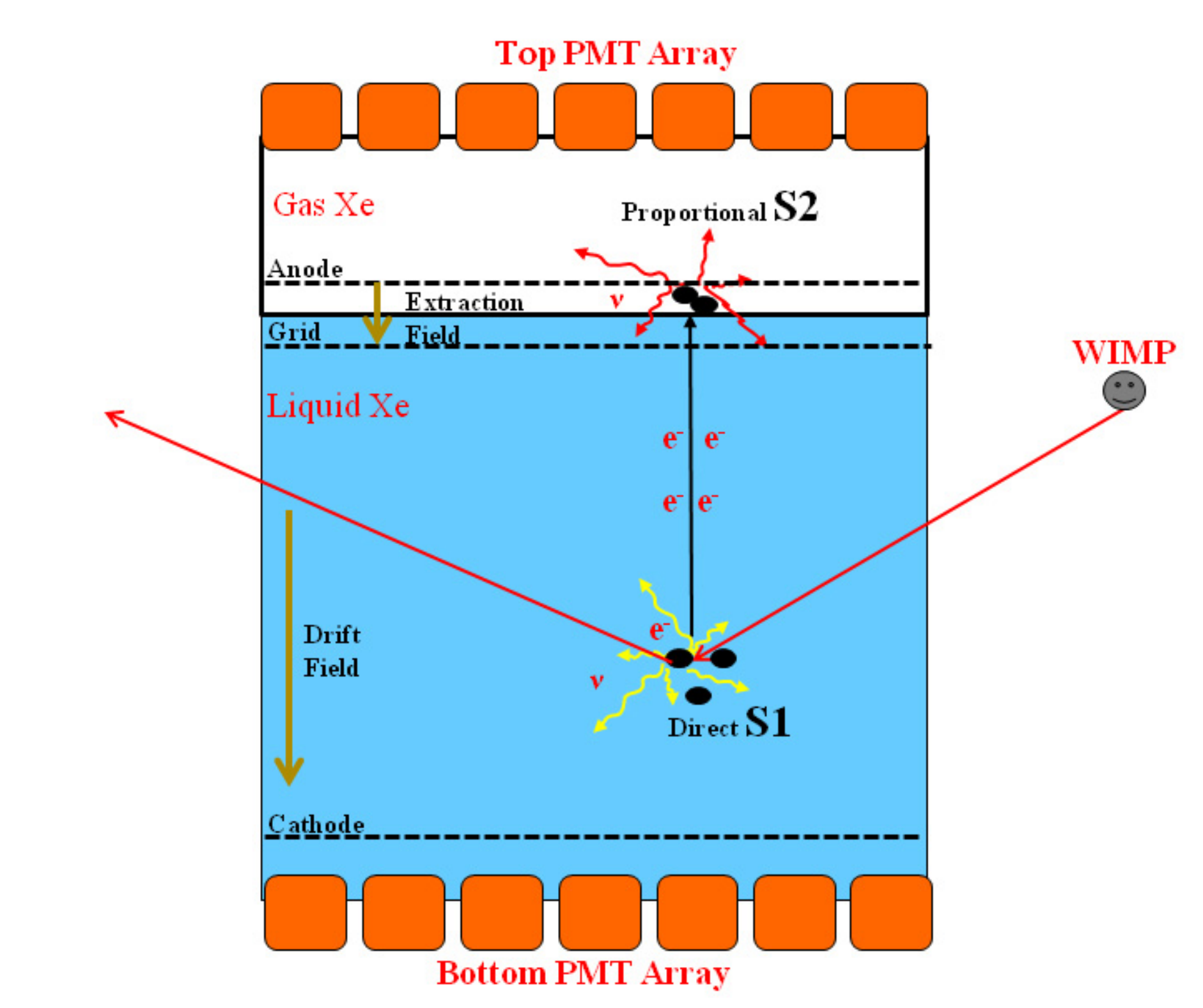}}
\caption{Working principle of a two-phase xenon TPC.}
\label{figure 1}
\end{figure}

The design of the TPC for a sensitive experiment is driven by the requirements of
maximizing light detection while maintaining the low background necessary for a rare event detection.
Due to the effect of total internal reflection at the liquid-gas interface, majority of the S1 light is detected by the bottom PMTs.
In XENON100 detector, about 80\% of the S1 light is detected by the bottom PMTs \cite{bib12}.
For the two detector configurations studied here, our simulation (see section 4) shows that about 67\% to 93\% of the S1 light, depending on the input parameters and detector configurations, is detected by the bottom PMT array.
For a ton-scale detector, the 3-inch-diameter R11410 PMT, as shown in figure 2 (left),
is a good candidate for the bottom PMT array to optimize the S1 light collection efficiency while reducing the number of electronics channels.
On the other hand, the R11410 PMT \cite{bib13} has a higher radioactivity and is much more expensive than the 1-inch-square R8520 PMT (figure 2, right).
As shown in table 1, the ${}^{60}\rm{Co}$ contamination per R11410 PMT is a factor of 3 $\sim$ 5 times that for a R8520 PMT.
Considering that the active area of R11410 is 7 times larger than that of R8520, the radioactivity per active area for the R11410 PMT is lower than that for a R8520 PMT.
Thus for a closely-packed configuration, such as for the bottom PMT array, R11410 is a better choice.
However, for the top PMT array, a loosely-packed array with the same number of R8520 PMTs provides similar performance in terms of light collection as that with R11410 PMTs, while reducing the total amount of radioactivity.

\begin{figure}[htd]
\centerline{\includegraphics[width=.6\textwidth]{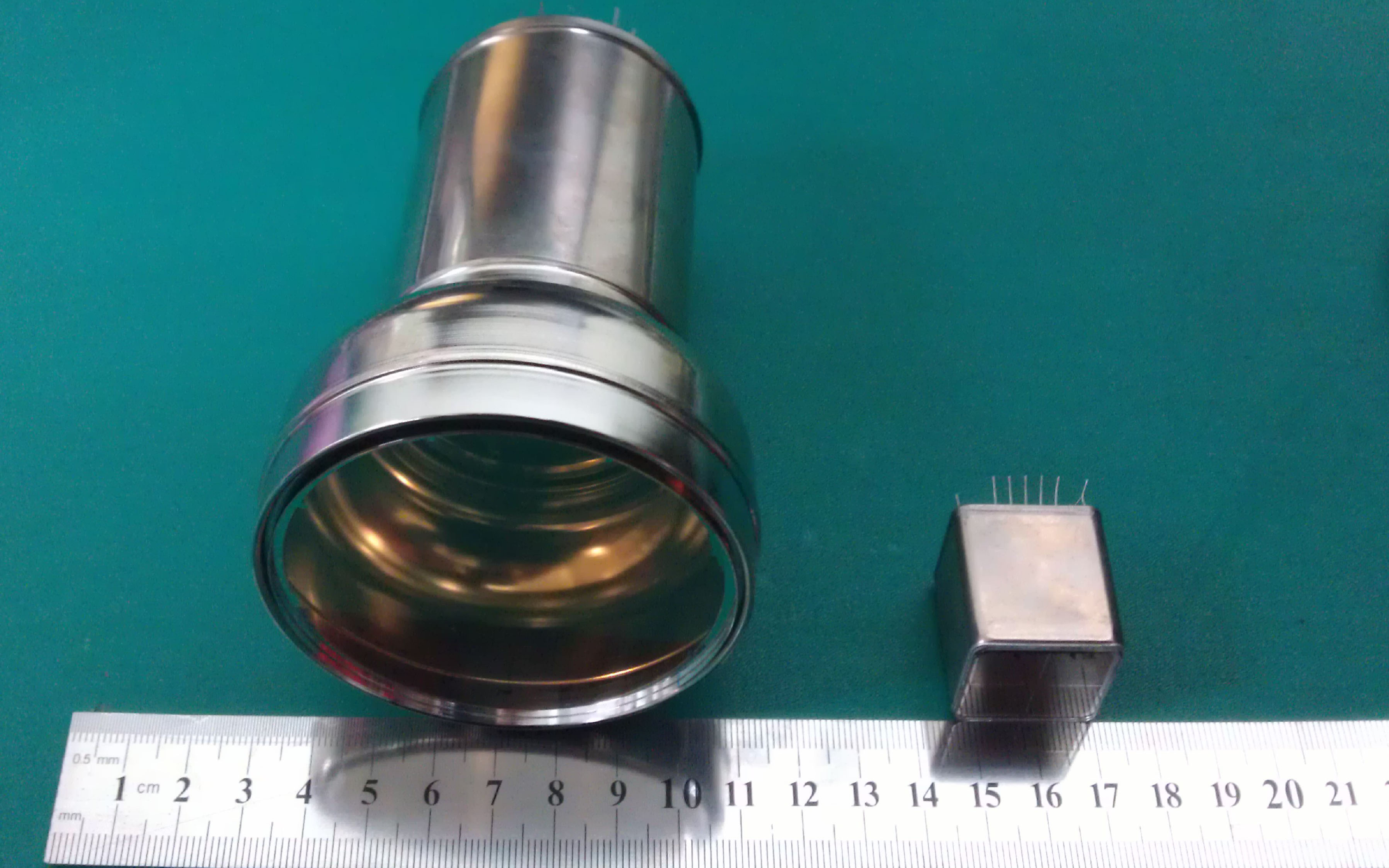}}
\caption{A photo of the 3-inch-diameter R11410 (left) and 1-inch-square R8520 (right) Hamamatsu PMTs.}
\label{figure 2}
\end{figure}

We studied two configurations for the TPC design for a liquid xenon detector with
a sensitive target of 1-m diameter and 1-m height, corresponding to a mass about 2.4 ton,
enclosed in the polytetrafluorethylene (PTFE) reflectors, as shown in figure 3.
For configuration 1, we maximize the S1 light collection efficiency by packing densely the R11410 PMTs for the bottom array.
The dead area between the PMTs is covered by PTFE reflectors.
A grounded screening mesh is placed above the bottom PMTs to shield them from the high voltage  on the cathode,
which is 5 cm above the screening mesh.
In total, there are 151 and 121 PMTs in the top and bottom arrays respectively.
The top array has 30 more PMTs compared to the bottom array for a good radial position reconstruction using
localized proportional S2 light \cite{bib12}.
In addition, there is a grid and an anode electrode on the top.
For configuration 2, there are 121 R11410 PMTs for the bottom array, which is the same for config.1.
For the top PMT array, we use the same number (151) of R8520 PMTs to replace the R11410 PMTs at the same
locations for config.1. Thus, the two configurations require same number of electronics channels,
while they are different in terms of background contribution from the top PMT array.
The light collection efficiency and the radial position sensitivity are also different.

\begin{figure}[htd]
\centerline{\includegraphics[width=.53\textwidth]{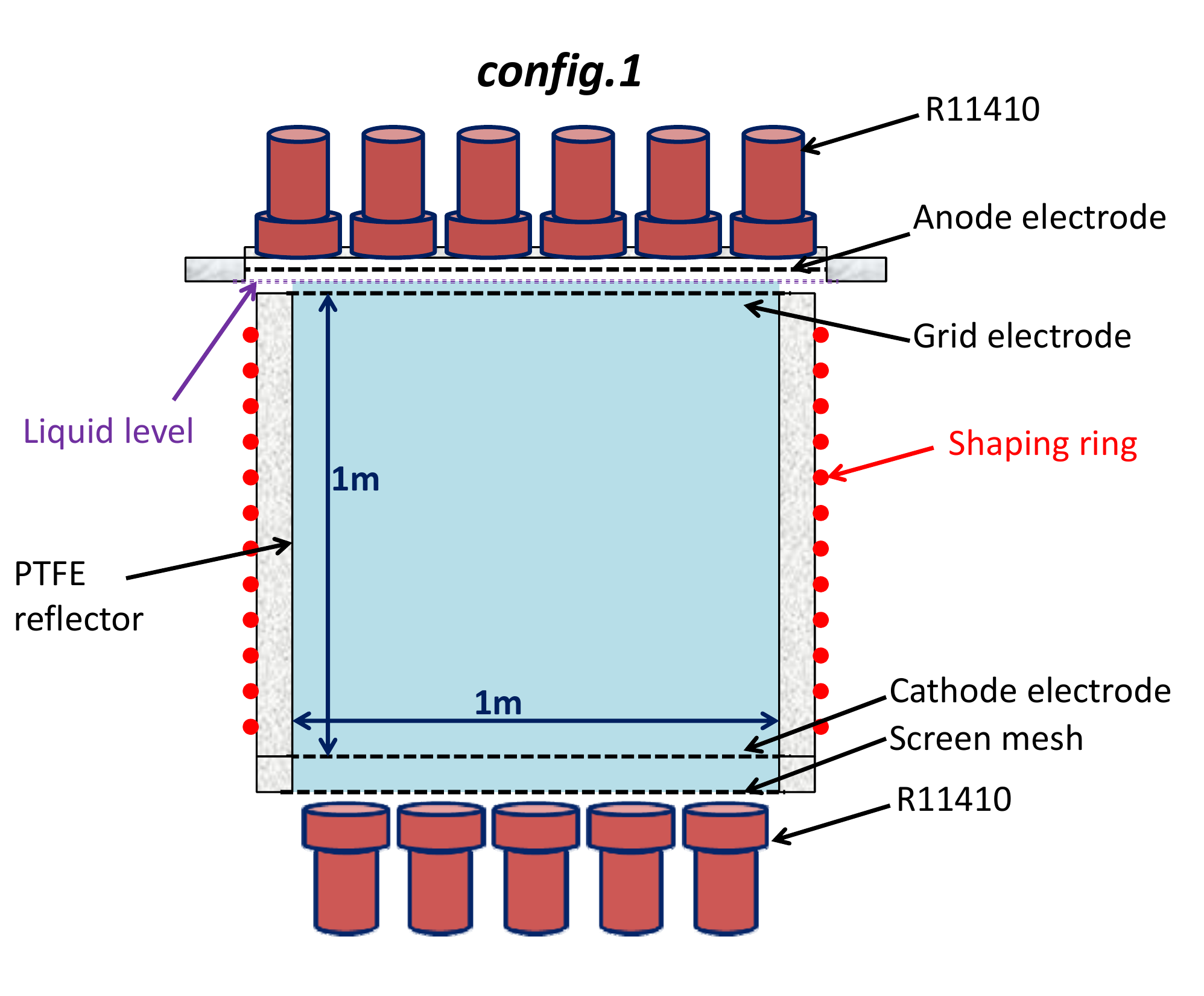} \hspace{1cm}
           \includegraphics[width=.41\textwidth]{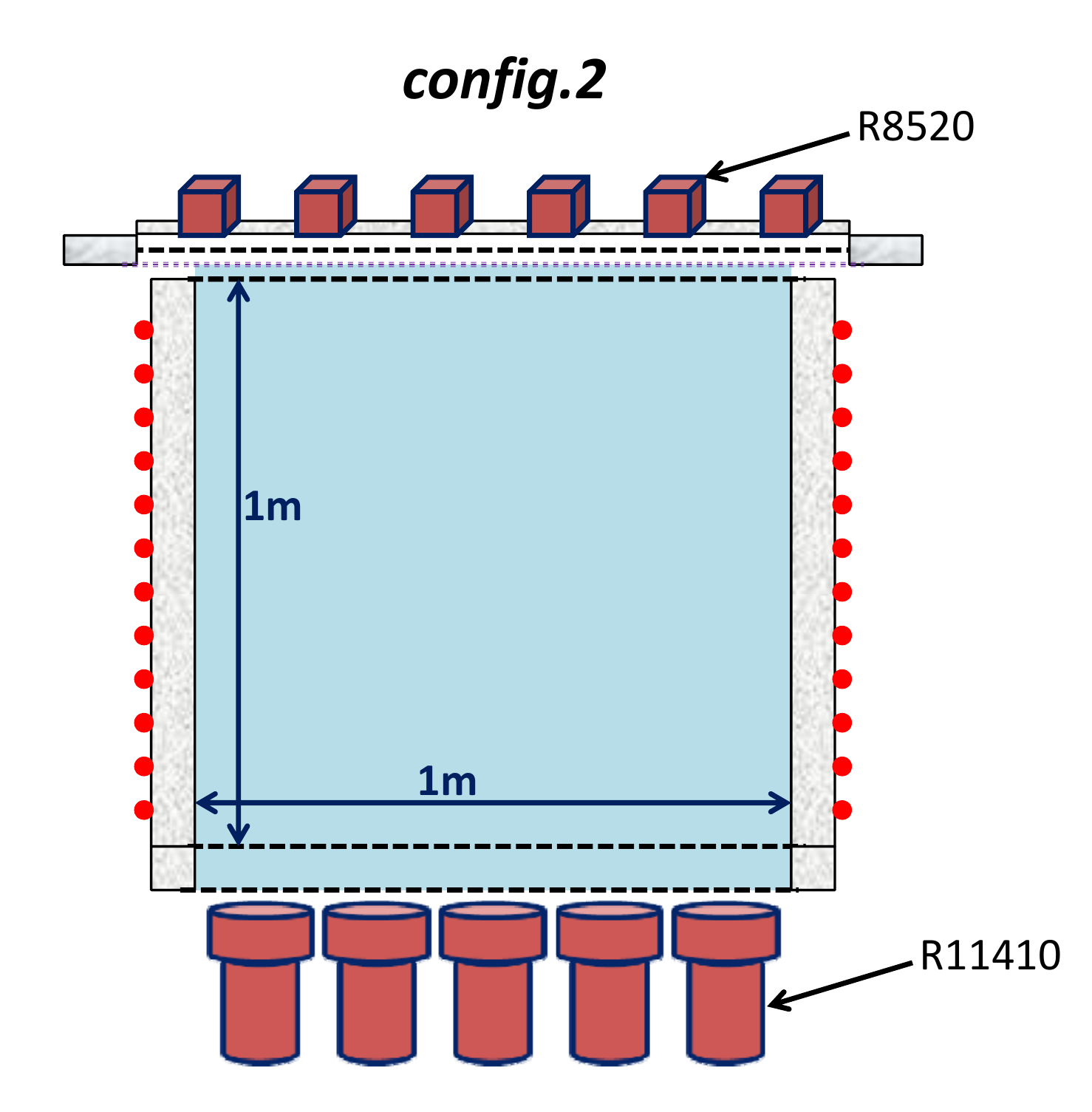}}

\centerline{\includegraphics[width=.26\textwidth]{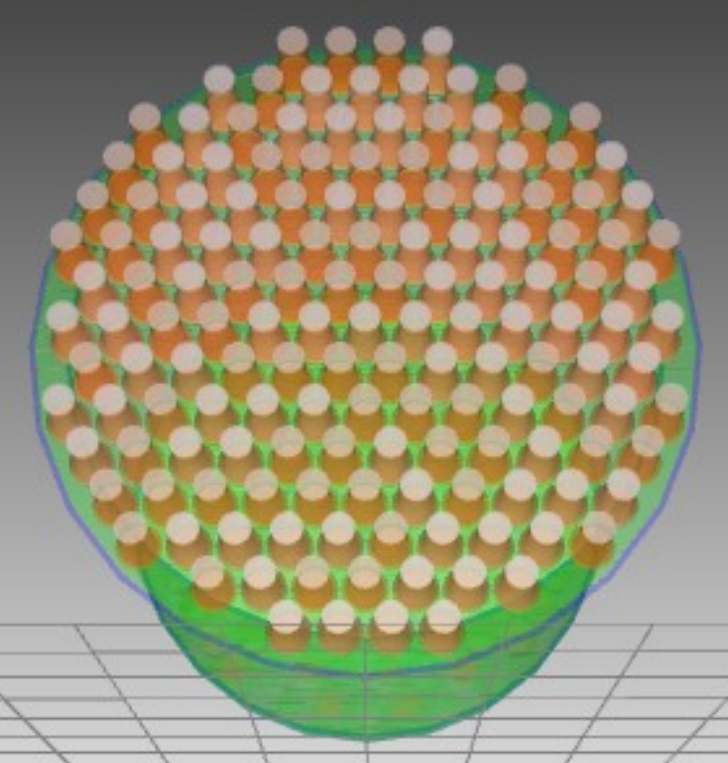} \hspace{3.5cm}
           \includegraphics[width=.275\textwidth]{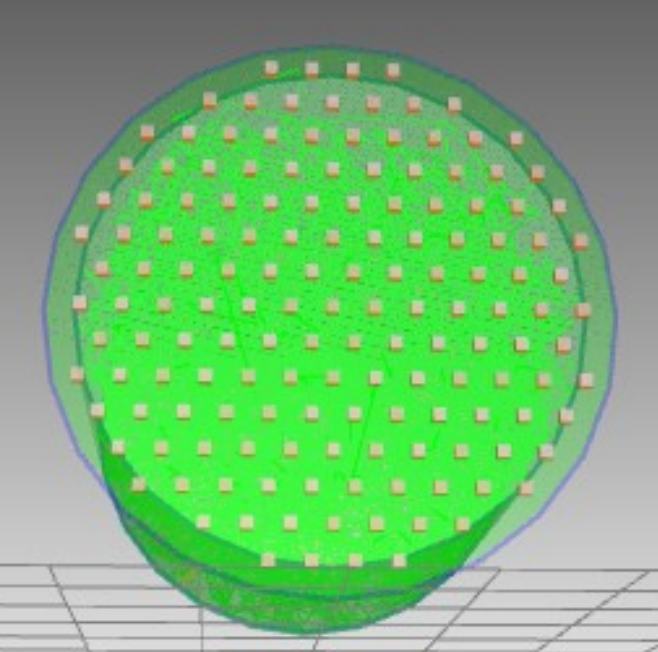}}
\caption{The conceptual design drawings of the two configurations for a ton-scale LXe detector.
         (Left) config.1: TPC structure with 151 R11410 PMTs on the top and 121 R11410 PMTs on the bottom. The figure below shows the arrangement of the 151 top R11410 PMTs.
         (Right) config.2: TPC structure with 151 R8520 PMTs on the top and 121 R11410 PMTs on the bottom. The figure below shows the arrangement of the 151 top R8520 PMTs.
         In addition to the PTFE reflectors on the side, there are PTFE reflectors in between the PMTs on both the top and the bottom arrays.}
\label{figure 3}
\end{figure}

\section{Background consideration}
For all experiments dealing with the rare event searches,
the reduction of the radioactive background is one of the most important and difficult tasks.
There are two types of background, electron recoils and nuclear recoils.
The electron recoil background is mostly from the gamma rays through the radioactive decay.
The nuclear recoil background is from neutrons from spontaneous fission, ($\alpha$, n) reactions and muon spallations \cite{bib14}.
The external gammas and neutrons from the muons and laboratory environment,
can be reduced by operating the detector at deep underground labs and by placing active or passive shield materials around the detector.

The radioactivity of the detector material also contributes to the background. Despite the numerous effort to reduce the radioactivity,
the PMTs are still the dominant internal background source in the detector material due to their close distance to the LXe target.
Other detector materials, such as the vessel containing the liquid xenon, can be made of ultra-pure materials,
limiting their background contribution to be sub-dominant compared to the PMTs.
For example, in XENON100 detector, the background is dominated by PMTs\cite{bib14c}.
The second major background from the 316Ti stainless steel vessel is about 40\% of that from PMTs. By using lower radioactive materials, such as  oxygen-free high conductive (OFHC) copper or pure titanium \cite{Titanium}, the background from the detector vessel can be further reduced.

For the radioactivity of the R8520 PMT, we adopt the screened values from XENON100 collaboration \cite{bib16}.
The average contamination levels of the main isotopes are listed in table 1.
For the radioactivity of R11410 PMTs, both XENON100 and LUX collaborations reported their results in \cite{bib13, bib16}.
We adopt the screened values from LUX's results which have a better sensitivity.
Both electron recoil and nuclear recoil background from the PMTs are simulated according to the contamination levels in table 1 for the two configurations.

\begin{center}	
	\begin{table}
	\caption{Screening results for different batches of PMTs as reported by different groups. The unit is mBq/PMT.}
	\centering
	\begin{tabular}{|c|c|c|c|c|c|}
	\hline
         PMT Type          	&	${}^{60}\rm{Co}$     &	${}^{40}\rm{K}$	     &	${}^{238}\rm{U}$     	 &	${}^{226}\rm{Ra}$    &	${}^{232}\rm{Th}$    \\ \hline	
	R8520 PMT \cite{bib16}  & 	0.75 $\pm$ 0.08 &	8.1 $\pm$ 0.9    &	0.25 $\pm$ 0.04      &   -     			&   0.5 $\pm$ 0.1   \\
	R11410 PMT\cite{bib13}  &	2.0  $\pm$ 0.2  & 	< 8.3	         &	< 6.0                &  < 0.4  			&   < 0.3           \\
	R11410 PMT\cite{bib16}  &	3.5  $\pm$ 0.6  & 	13 $\pm$ 4	     &	< 95                 &  < 2.4  			&   < 2.6           \\
	\hline
	\end{tabular}
	\label{tab1}
	\end{table}
\end{center}

\subsection{Electron recoil background}
We simulated $6 \times 10^{7}$ events for each isotope for both configurations, and calculated the electron recoil background rates with the upper
and the lower limits of the radioactivity for each isotope in the following analysis.
If a fixed radioactivity value is given, we use the mean value plus and minus the error as the upper and the lower limits.
If only an upper limit is given, we set the lower limit to zero.
In figure 4, we show the spatial distribution of the single-scatter electron recoil event rates for events below 50 $\rm{keV}_{ee}$ ($\rm{keV}_{ee}$ is the electron equivalent energy in keV) with
the upper limits of radioactivity for these two configurations.

\begin{figure}[htd]
\centerline{\includegraphics[width=.5\textwidth]{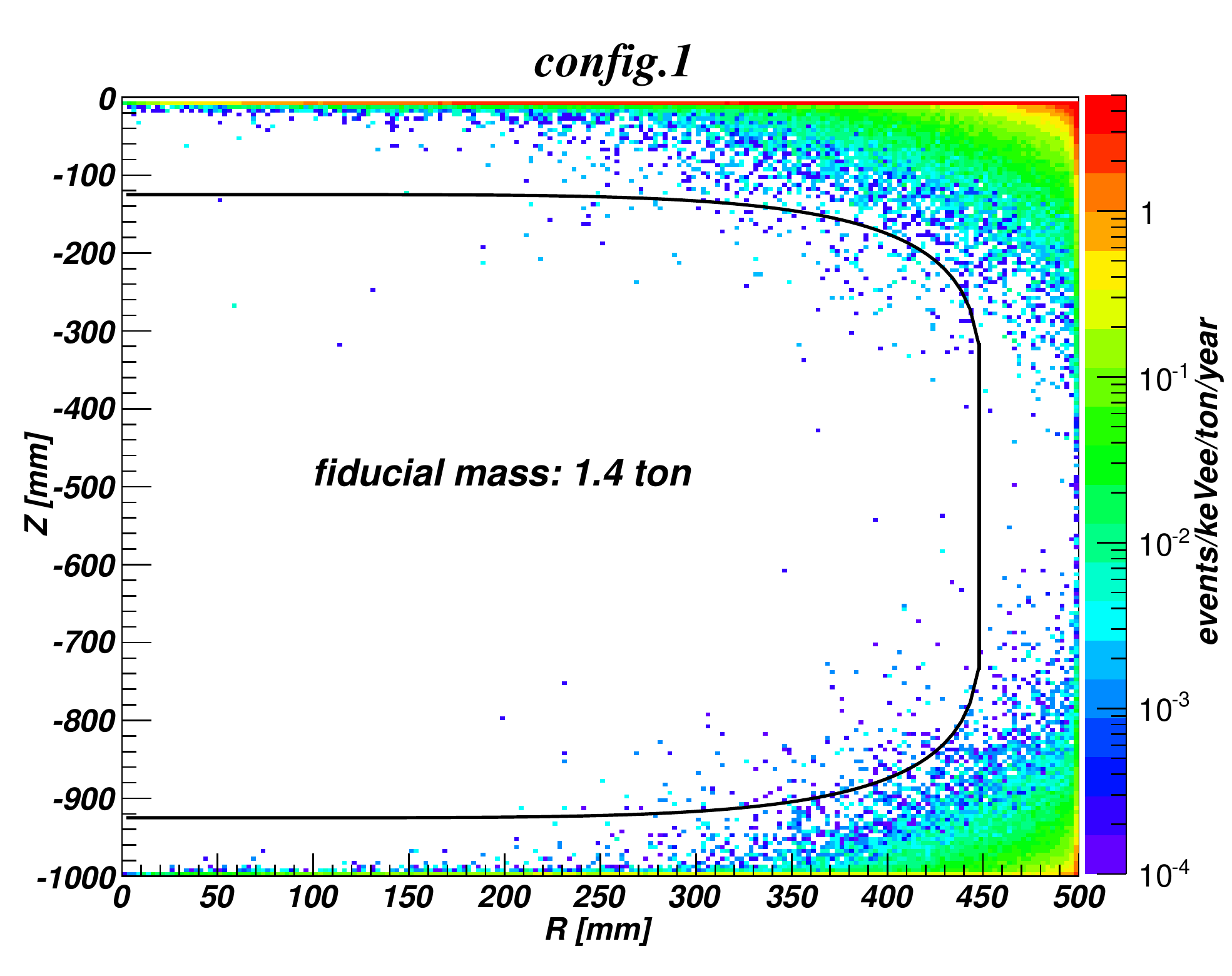} \hspace{1cm}
            \includegraphics[width=.5\textwidth]{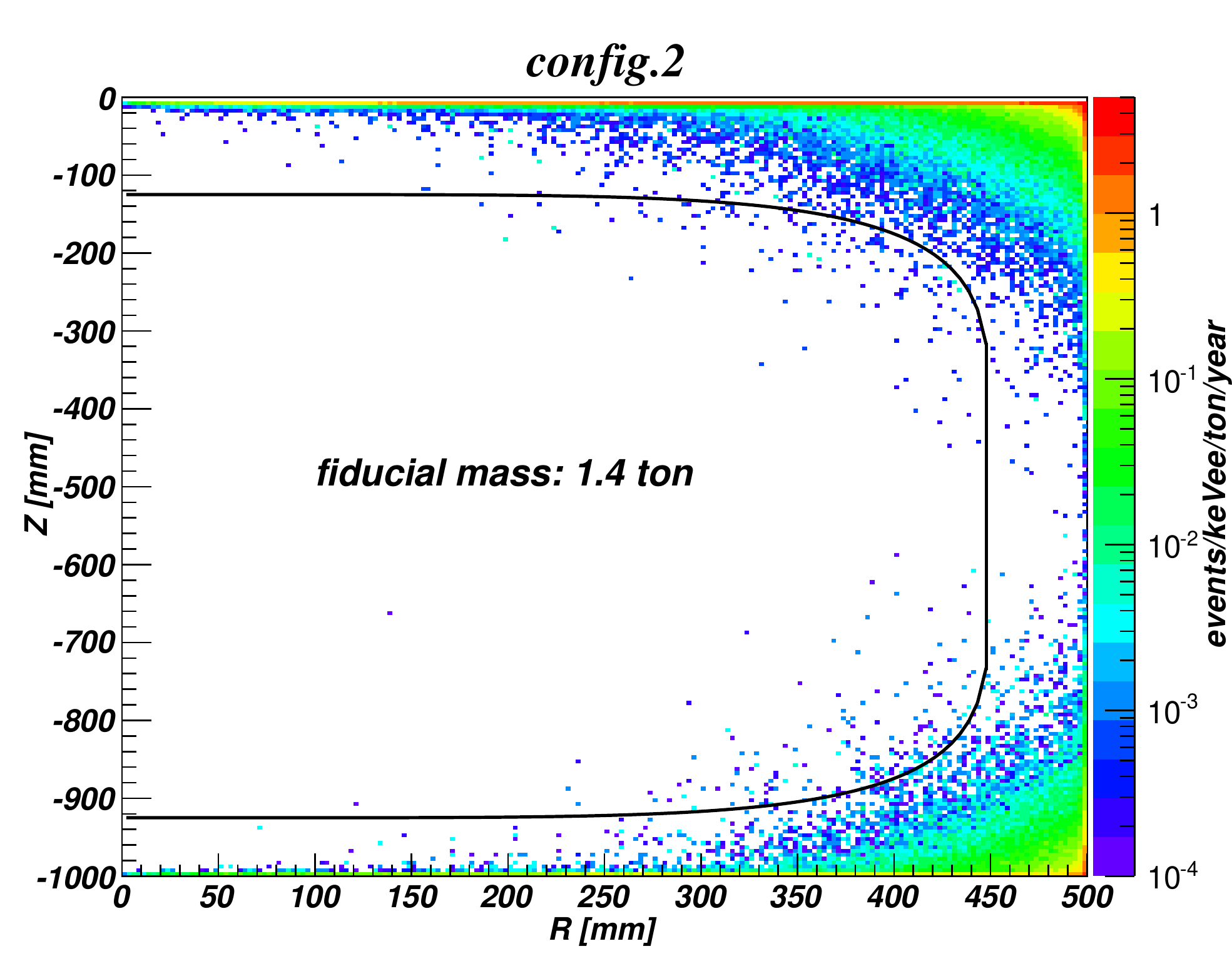}}
\caption{Spatial distributions of single-scatter electron recoil background event rates from the PMTs for events below 50 $\rm{keV}_{ee}$ for the two configurations.
		 The upper limits of radioactivity of each isotope as listed in table 1 are used.}
\label{figure 4}
\end{figure}

The presence of electron recoil background events will reduce the sensitivity of a detector.
To minimize the background, fiducial volume cut needs to be applied.
Since the electron recoil background events from PMTs are mostly located at the two corners on the top and bottom at the large radial positions,
we choose a fiducial volume in a superellipsoid shape according to the following function:
\begin{equation}
(\frac{r}{450})^{5.5} + (\frac{z + 525}{400})^{5.5} \leq 1.
\end{equation}
We limit the radial position to be less than 450 mm to minimize the background contribution from the PTFE wall and the detector vessel.
The $z$ position is chosen between -125 mm and -925 mm such that the contribution from the top and bottom PMTs are roughly equal.
The fiducial volume is slightly shifted towards the bottom due to the additional 5 cm self-shielding LXe between the cathode and the bottom PMT array.
The fiducial volume selection provides a target mass of 1.4 ton.
The simulated electron recoil background event rates in the fiducial volume in these two configurations are shown in table 2.

\begin{center}
	\begin{table}[hbt]
	\caption{Simulated single-scatter electron recoil background event rates from the PMTs, for events below
	          50 $\rm{keV}_{ee}$ after the fiducial mass selection. The unit is events/$\rm{keV}_{ee}$/ton/year.
	        The statistical error in the simulation is about 5\%. The two values correspond to the lower and the upper limits of the radioactivities listed in table 1.}
	\centering
	\begin{tabular}{|c|c|c|c|}
	\hline
       Configuration Type     &	Top PMTs        &	Bottom PMTs	 &	All PMTs     \\ \hline	
		config.1  			  & 0.1 - 0.4      &	0.1 - 0.3   &	0.2 - 0.8  \\
		config.2  			  &	0.1 - 0.2      & 	0.1 - 0.3	 &	0.3 - 0.5  \\
	\hline
	\end{tabular}
	\label{tab2}
	\end{table}
\end{center}

We found, in either configuration, that the electron recoil background from the PMTs is below 1 $\rm{events/keV_{ee}/ton/year}$ (or $2.7\times10^{-6}$ $\rm{events/keV_{ee}/kg/day}$).
At such a low background level, the background from the beta decays of ${}^{85}\rm{Kr}$,
the double beta decays of ${}^{136}\rm{Xe}$ and the electron scatterings of the $pp$ solar neutrinos become dominant.
They produce electron recoil background uniformly distributed in the fiducial volume.
The beta decays of ${}^{85}\rm{Kr}$ from one part-per-trillion (ppt) Kr/Xe contamination in xenon will contribute $2\times10^{-5}$ $\rm{events/keV_{ee}/kg/day}$ electron recoil events
in the target \cite{bib12}. In order to reduce the contribution from beta decays of ${}^{85}\rm{Kr}$ to be sub-dominant compared to the PMTs,
the Kr/Xe contamination needs to be lowered below 0.01 ppt ($0.2\times10^{-6}$ $\rm{events/keV_{ee}/kg/day}$) with the further development of the Kr removal technology \cite{bib14a}.
The double beta decays of ${}^{136}\rm{Xe}$ in the natural xenon will contribute an electron recoil background
at the level of $0.7 E \times 10^{-6}$ $\rm{events/keV_{ee}/kg/day}$ \cite{bib14b}, where $E$ is the energy in $\rm keV_{ee}$,
at the low energy region relevant to the dark matter searches. This background can only be removed by using ${}^{136}\rm{Xe}$-depleted xenon.
More seriously, the background from electron scatterings of the $pp$ solar neutrinos is estimated at the level of $1.4 \times 10^{-5}$ $\rm{events/keV_{ee}/kg/day}$ \cite{bib14b},
which is above the background from the PMTs in our simulation. This background can't be removed and is the dominant background for the ton scale xenon detector for dark matter searches.
Further reduction of the electron recoil background can only be achieved by the discrimination of electron/nuclear recoils based on the S2/S1 ratio.

\subsection{Nuclear recoil background}
To estimate the nuclear recoil background, we first calculated the neutron spectra and production rates from the spontaneous fission and ($\alpha$, n) in the PMTs due to the contamination of
${}^{238}\rm{U}$ and ${}^{232}\rm{Th}$ by the modified SOURCE4A code \cite{SOURCE}.
Assuming the radioactivity is uniformly distributed in the PMT's casing, window and photocathode,
we simulated $10^{6}$ neutron events for each part of the PMT using the generated neutron spectra. We then calculated the single-scatter nuclear recoil event rates based on the upper and lower limits of the ${}^{238}\rm{U}$ and ${}^{232}\rm{Th}$.
The simulated single-scatter nuclear recoil event rates in the fiducial volume for these two configurations are shown in table~3.
We found, in either configuration, the single-scatter nuclear recoil event rate is in the order of $10^{-3}~\rm{events/keV_{nr}/ton/year}$ ($\rm keV_{nr}$ is the nuclear recoil energy in keV),
almost three orders of magnitude lower than the single-scatter electron recoil event rate from PMTs.
For a 99.9\% electron recoil rejection power based on the S2/S1 ratio, we expect similar nuclear recoil and electron recoil background events in the WIMP search region from the PMTs.
However, they are still sub-dominant compared with the electron recoil background from the $pp$ solar neutrinos.

\begin{center}
	\begin{table}[hbt]
	\caption{Simulated single-scatter nuclear recoil background event rates from PMTs, for events below
	        50 $\rm{keV}_{nr}$ after the fiducial mass selection. The unit is events/$\rm{keV}_{nr}$/ton/year.
	        The statistical error in the simulation is about 1\%. The two values correspond to the lower and the upper limits of the radioactivities listed in table 1.}
	\centering
	\begin{tabular}{|c|c|c|c|}
	\hline
       Configuration Type     &	Top PMTs       					 				&	Bottom PMTs	 				&	All PMTs     \\ \hline	
		config.1  			  & 0 - $5.0\times 10^{-4}$          				&	0 - $1.0 \times 10^{-3}$    &	0 - $1.5 \times 10^{-3}$  \\
		config.2  			  &	$1.8 \times 10^{-5}$ - $2.6 \times 10^{-5}$     & 	0 - $1.0 \times 10^{-3}$	&	$1.8 \times 10^{-5}$ - $1.0 \times 10^{-3}$  \\
	\hline
	\end{tabular}
	\label{tab3}
	\end{table}
\end{center}

The other detector components can also contribute to the total nuclear recoil background.
In addition, muons can also induce neutron flux in rock and shielding materials.
The simulation of these additional neutron background is beyond the scope of this study.
Based on a study of simulating these neutron backgrounds~\cite{SOURCE},
we take a conservative estimation by assuming the other sources contribute a $3 \times 10^{-3}~\rm{{keV}_{nr}/ton/year}$ of neutron background,
which is about two to three times of the upper limits of the neutron background rates from the PMTs.
We include the total nuclear recoil background in the sensitivity calculation in section 6.

\section{Primary scintillation light collection efficiency}
One of the challenges of a two-phase xenon detector is to efficiently detect the small number of S1 photons
associated with the low energy nuclear recoils from a WIMP scattering.
We simulate the light collection efficiency (LCE) by considering several key properties, including the PTFE reflectivity and the LXe's absorption length.

PTFE is known to have a high reflectivity and often used as reflectors to improve the light collection efficiency of scintillation photons in a LXe detector.
But the reflectivity of PTFE also depends on the surface treatment and thus varies from sample to sample. In our light collection simulation, we vary the PTFE reflectivity from 85\% to 95\%.
Some studies show that the reflectivity of PTFE in GXe could be 10\% to 25\% smaller than that in LXe \cite{Claudio F. P. Silva}.
We checked its effect on the light collection by setting the reflectivity of PTFE in GXe from 60\% to 70\%.
The impact on the light collection is less than 1\%. This is because only small fraction of photons entering into the gas xenon due to the total internal reflection at the liquid-gas interface.
The reflectivity of PTFE between the top PMTs thus plays a less important role for the total light collection.

The absorption length ($\lambda_{abs}$) of the UV light in LXe depends on the purity level \cite{Baldini}.
The main impurities are water and oxygen \cite{bib17}.
The light is mostly attenuated by water, whereas for the charge the most harmful impurity is oxygen.
The purity can be improved by  constantly recirculating xenon gas through a high temperature getter,
which removes impurities by chemically bonding them to the getter material.
The XENON100 collaboration has demonstrated sub-ppb (part-per-billion) of oxygen-equivalent purity level with an electron lifetime of more than $0.5~\rm{ms}$~\cite{bib12}.
We assume that the water impurity in xenon can be achieved in a similar level.
According to~\cite{Baldini}, $\lambda_{abs}$ can reach 1 m with a 100 ppb  water contamination in xenon.
Thus a 10 ppb water contamination in xenon will allow $\lambda_{abs}$  to be at about 10 m.
We used $\lambda_{abs}$ from 2.5 to 10 m for the input for the detector simulation.

The Rayleigh scattering length ($\lambda_{sca}$) will deflect the direction of the light, thus increasing the total length for a photon before it reaches the PMTs.
The $\lambda_{sca}$ value was measured between 29 and 50 cm \cite{bib18}. We adopt a value of 40 cm in our simulation.
The electrodes will also block part of the light.
In our simulation, the electrode is composed by stainless steel wires with 200 $\mu$m diameter and 5 mm spacing between neighboring ones.
The transparency of each electrode is about 95\%.

\begin{figure}[htd]
\centerline{\includegraphics[width=.5\textwidth]{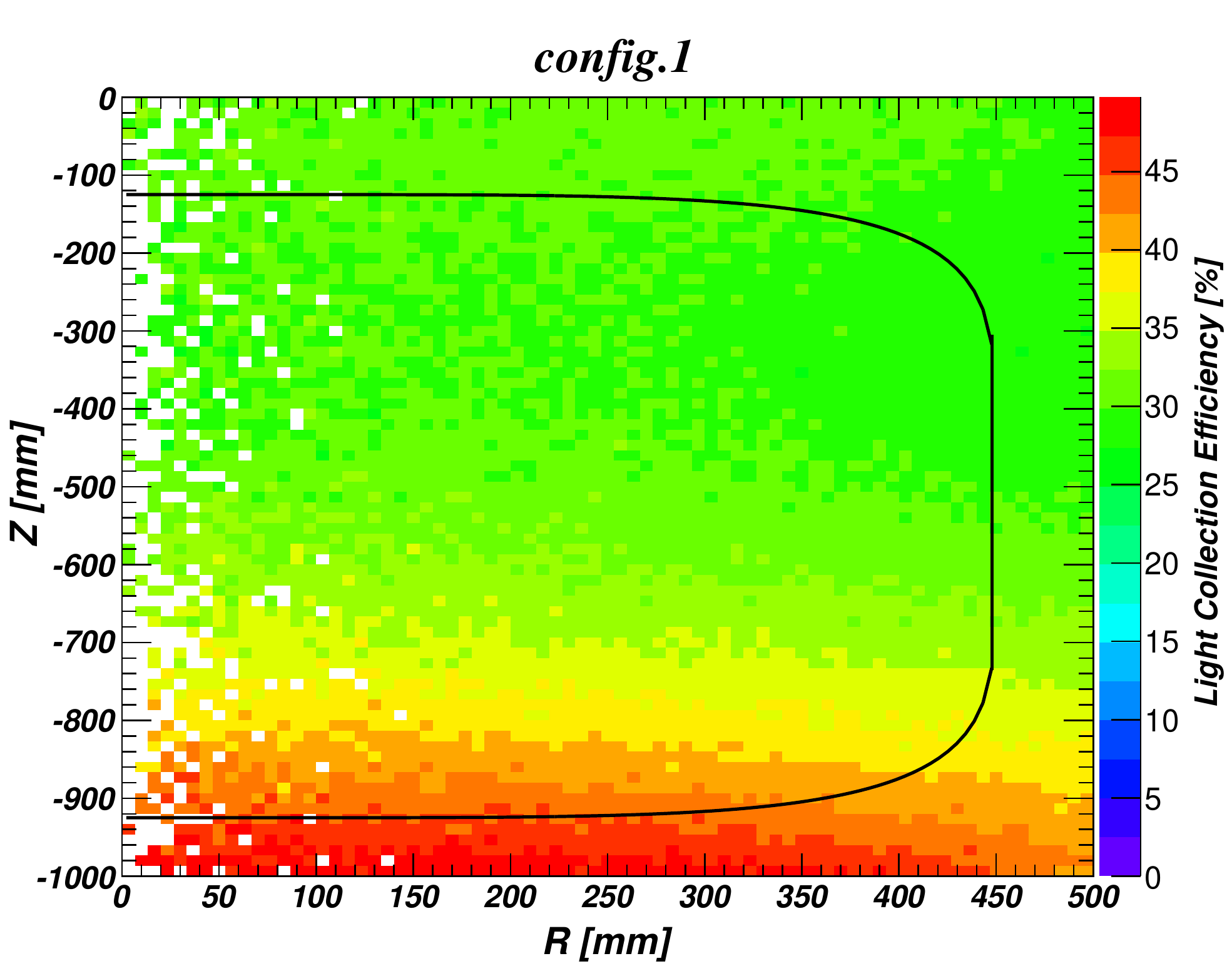} \hspace{1cm}
            \includegraphics[width=.5\textwidth]{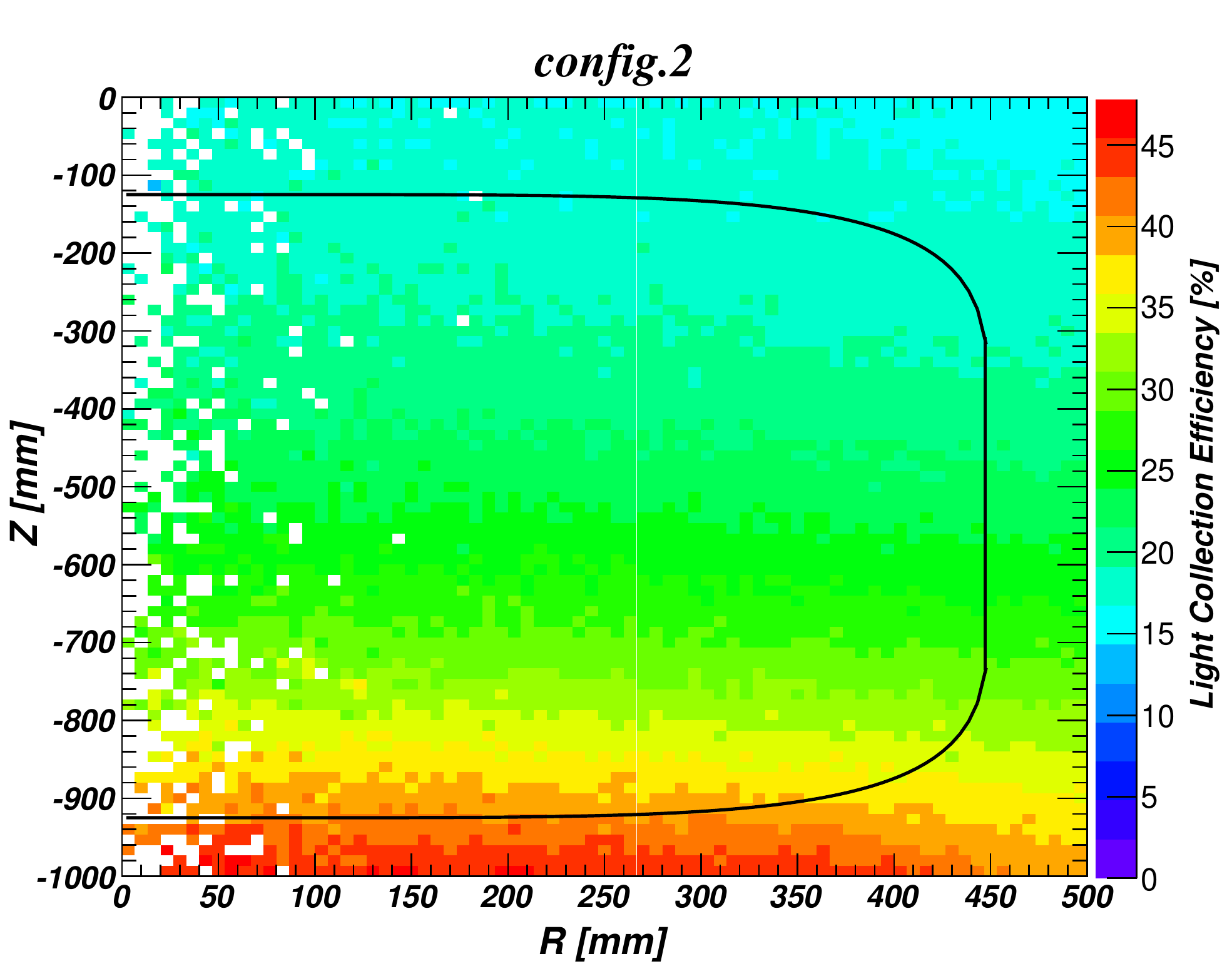}}
\caption{S1 LCE distribution through-out the active volume of TPC with PTFE reflectivity,
			$\lambda_{abs}$ and $\lambda_{sca}$ at 95\%, 5 m and 40 cm, respectively.}
\label{figure 5}
\end{figure}

The S1 LCE is defined as the ratio between the total number of photons
collected by the photocathode of all the PMTs and the total number of scintillation photons produced in LXe from one event.
The events are randomly generated and event positions are uniformly distributed in the LXe target during the simulation. The simulated LCE distribution through-out the LXe target is shown in figure 5.
The LCE is very similar for the bottom part of the TPC for the two configurations. But for the top part, the LCE in config.2 is slightly lower than that in config.1.
For the superellipsoid fiducial region, the distributions of LCE are shown in figure 6 (left), with the lowest,
the median and the highest LCE respectively for both configurations.
We summarize our simulation results in figure 6 (right) with different input parameters.
Although LCE for the detector with config.1 is about 1.3 times that for config.2,
the main parameters affecting the LCE are the PTFE reflectivity and the absorption length.
The highest LCE is almost a factor of 2.5 times the lowest one. Thus selecting a highly reflective PTFE and effective xenon purification are the key factors to achieve a high LCE.

\begin{figure}[htd]
\centerline{\includegraphics[width=.5\textwidth]{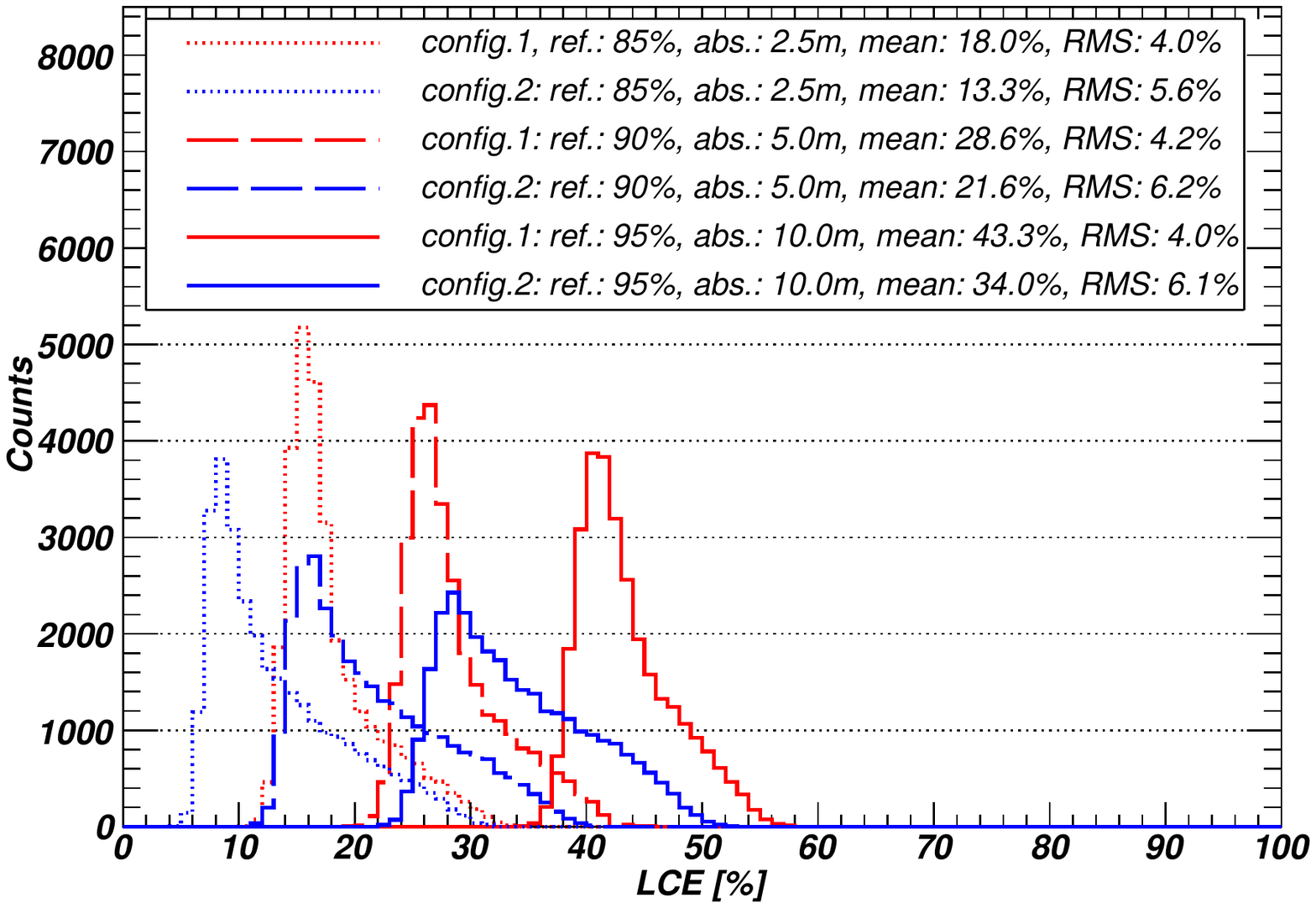} \hspace{1cm}
            \includegraphics[width=.5\textwidth]{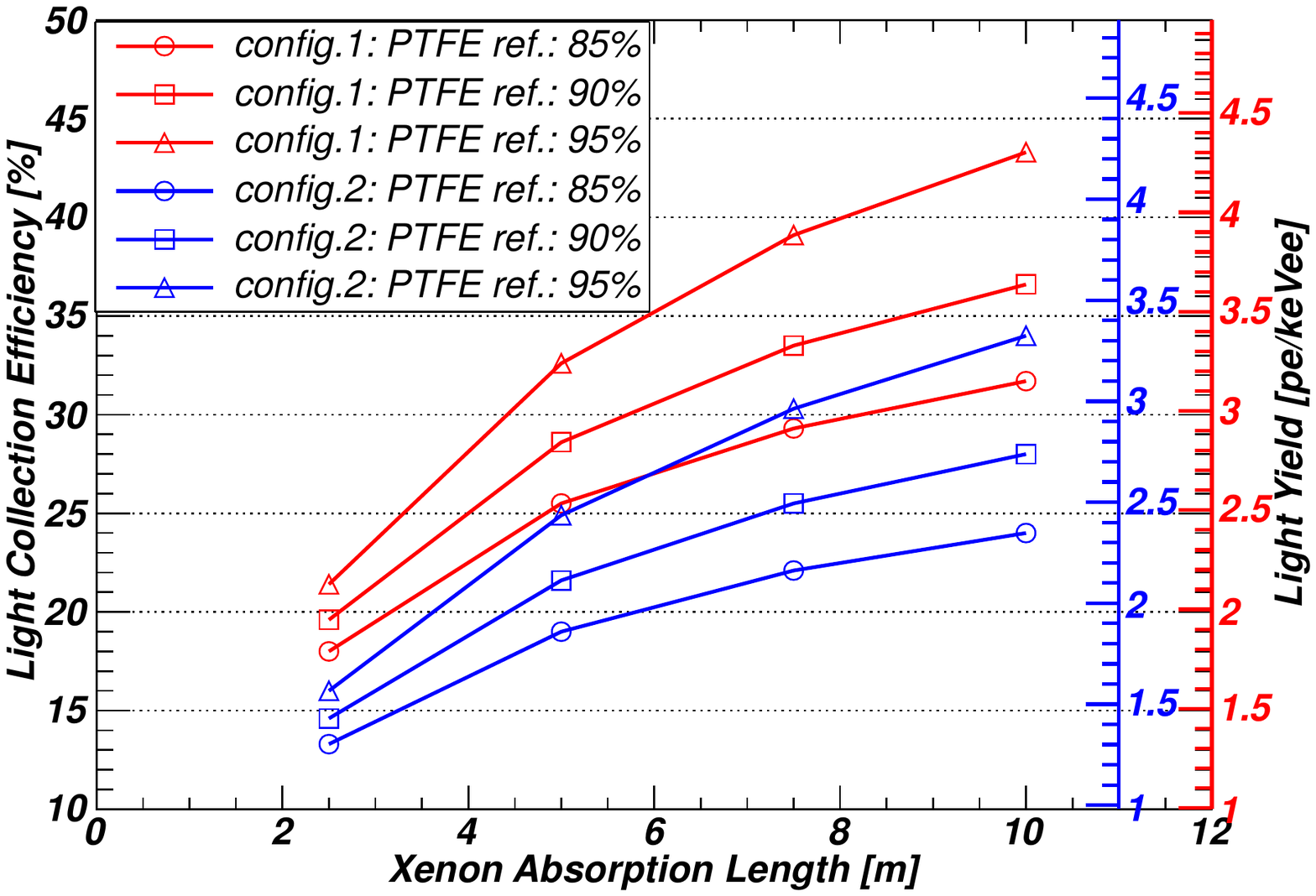}}
\caption{(Left) S1 LCE distributions in the fiducial volume for config.1 and config.2 with different PTFE reflectivities (ref.) and LXe absorption lengths (abs.) respectively.
                The mean and variation (RMS) of the LCE values are shown in the figure legend.
		 (Right)  S1 LCE and the corresponding light yield after the fiducial selection for different PTFE reflectivities
		          and absorption lengths.}
\label{figure 6}
\end{figure}

We also calculate the light yield ($LY$) in unit of  photoelectrons/$\rm{keV_{ee}}$($\rm{pe/keV_{ee}}$) according to equation (4.1) at a nominal drift field of 1 kV/cm.
The quantum efficiency ($QE$) of the PMTs is at least 30\% at room temperature \cite{hamamatsu,bib19} and we use an average $QE$ of 33\% for all tubes.
In the liquid xenon temperature, the $QE$ was measured to increase by $\sim5-11$\%~\cite{arxiv:1207.5432}.
The collection efficiency of photoelectrons from the photocathode to the first dynode is about 85\% for R11410 PMTs and 70\% for R8520 PMTs~\cite{hamamatsu}.
By taking the above factors, we use an effective quantum efficiency ($QE_{eff}$) of 31\% for R11410 PMTs and 25\% for R8520 PMTs.
A $W_{ph}$ of 15.6 eV for 122~keV gamma rays under zero electric field in LXe is adopted according to the NEST model \cite{bib20}.
The field quenching factor of light ($q_{f}$) is 50\% at a drift field of 1 kV/cm \cite{bib21}.

\begin{equation}
LY = \frac{1 \rm{keV_{ee}}}{W_{ph}} \times LCE \times QE_{eff} \times q_{f}.
\end{equation}

\section{Position sensitivity}

One advantage of the two-phase TPC technique for rare event detection is the possibility to precisely determine all three coordinates of an interaction vertex
in the target volume, which allows position dependent signal corrections and fiducial volume selection for background suppression.
The $z$-coordinate can be determined by the drift time between the prompt S1 and the delayed S2 signals.
The determination of the ($x, y$) position is realized by using the S2 signal distribution on the top PMT array. Several methods can be
used to reconstruct the position such as the minimum $\chi^{2}$, support vector machine (SVN), neural network (NN) \cite{bib12},
maximum likelihood and least squares methods \cite{ZEPLIN-III}. We use the NN method for this work.

We simulated $10^{6}$ S2 events with $10^{3}$ photons/event, which are uniformly distributed in the S2 production area for each configuration.
They are fed into the NN for training. Another set of S2 events uniformly distributed in the volume were randomly generated as ``the experimental data''
to test the position reconstruction for the two configurations.

\begin{figure}[htb]
\centerline{\includegraphics[width=.5\textwidth]{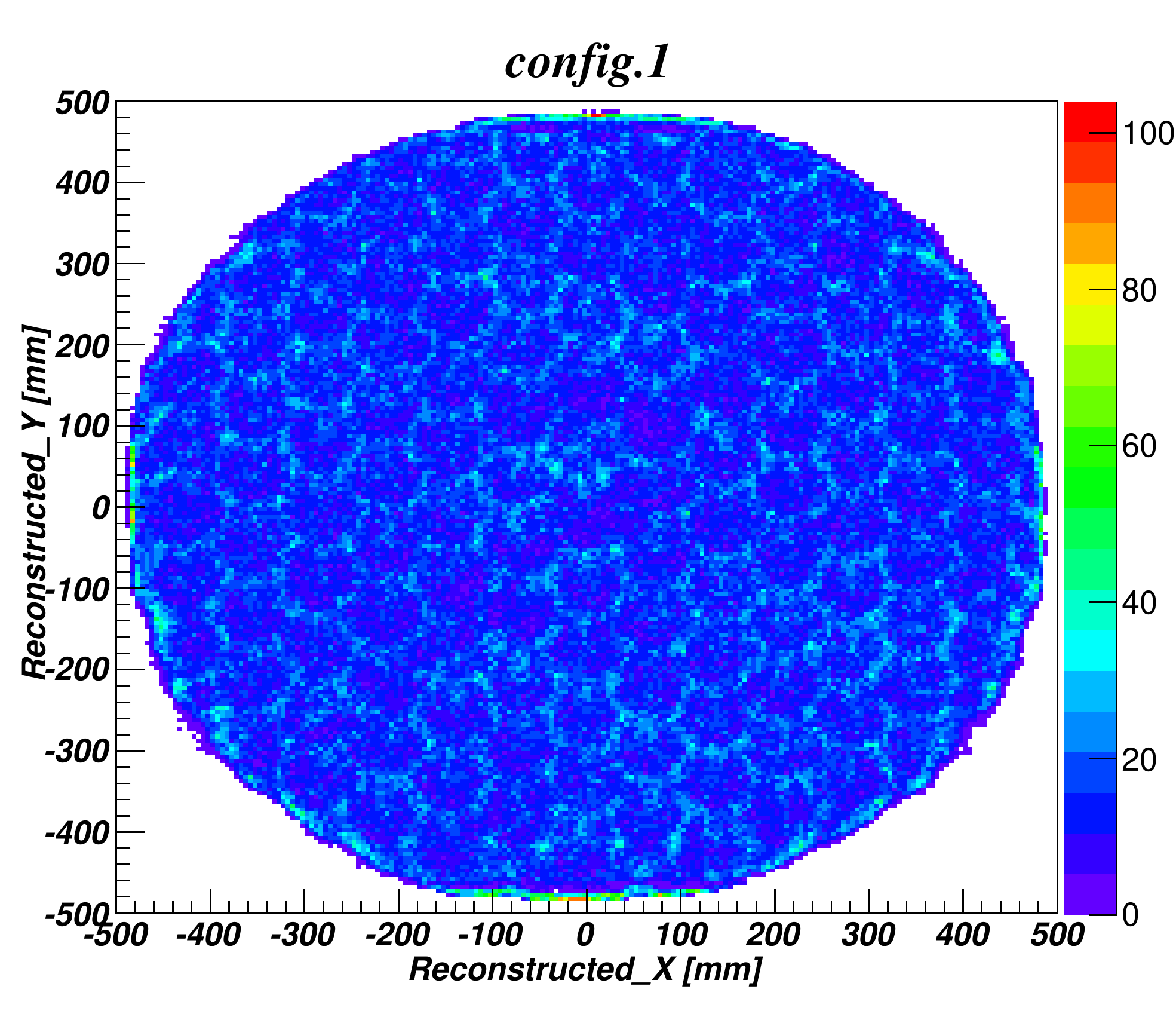} \hspace{1cm}
            \includegraphics[width=.5\textwidth]{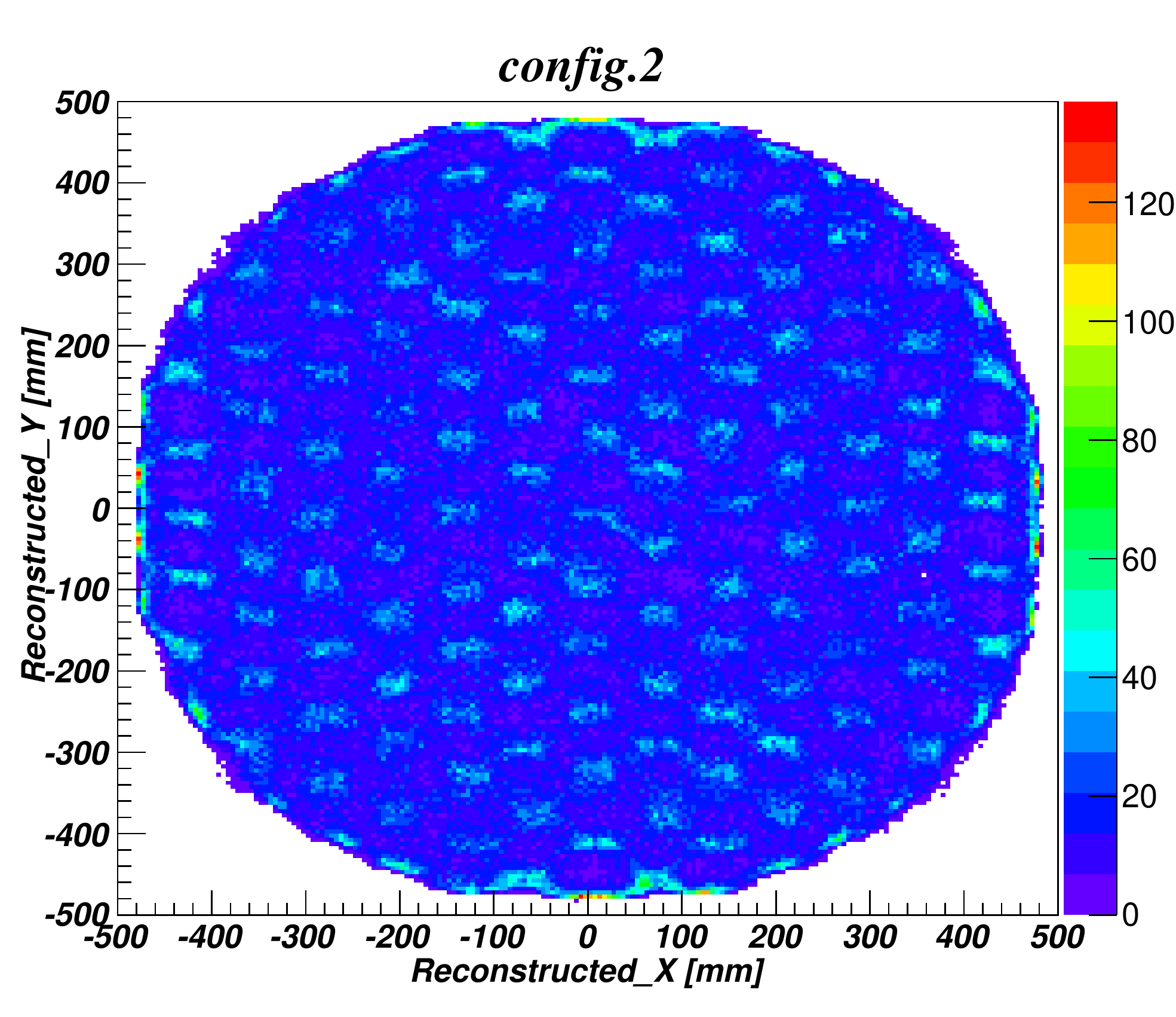}}
\caption{Event distributions after NN reconstruction for the two configurations.}
\label{figure 7}
\end{figure}

One main function of the position reconstruction is to suppress the background at the edge of the TPC.
We define the reconstruction error ($\triangle_{R}$) as the quality of the position reconstruction:
\begin{equation}
\triangle _{R} = R_{reconst} - R_{ori}
\end{equation}
$R_{reconst}$ and $R_{ori}$ stand for the reconstructed and the original radial position for each event respectively.
The distribution of reconstructed events is shown in figure 7 for both configurations.
For the top PMT array in config.1, more events are reconstructed in the gaps between PMTs while in config.2, more
events are reconstructed near the center of each PMT.
The main reason is the different space between neighboring PMTs, which for config.2 is much more separated than for config.1.
As shown in figure 8, the average reconstruction errors ($\triangle_{R}$) for the events in the fiducial volume are 0.1 mm and 0.9 mm for config.1 and config.2 respectively.
The Root-Mean-Square (RMS) values of $\triangle_{R}$, also referred as the reconstruction position resolutions,
are 5.4 mm and 10.4 mm for config.1 and config.2 respectively. Config.1 performs better due to a larger PMT coverage than config.2.

\begin{figure}[hbt]
\centerline{\includegraphics[width=.5\textwidth]{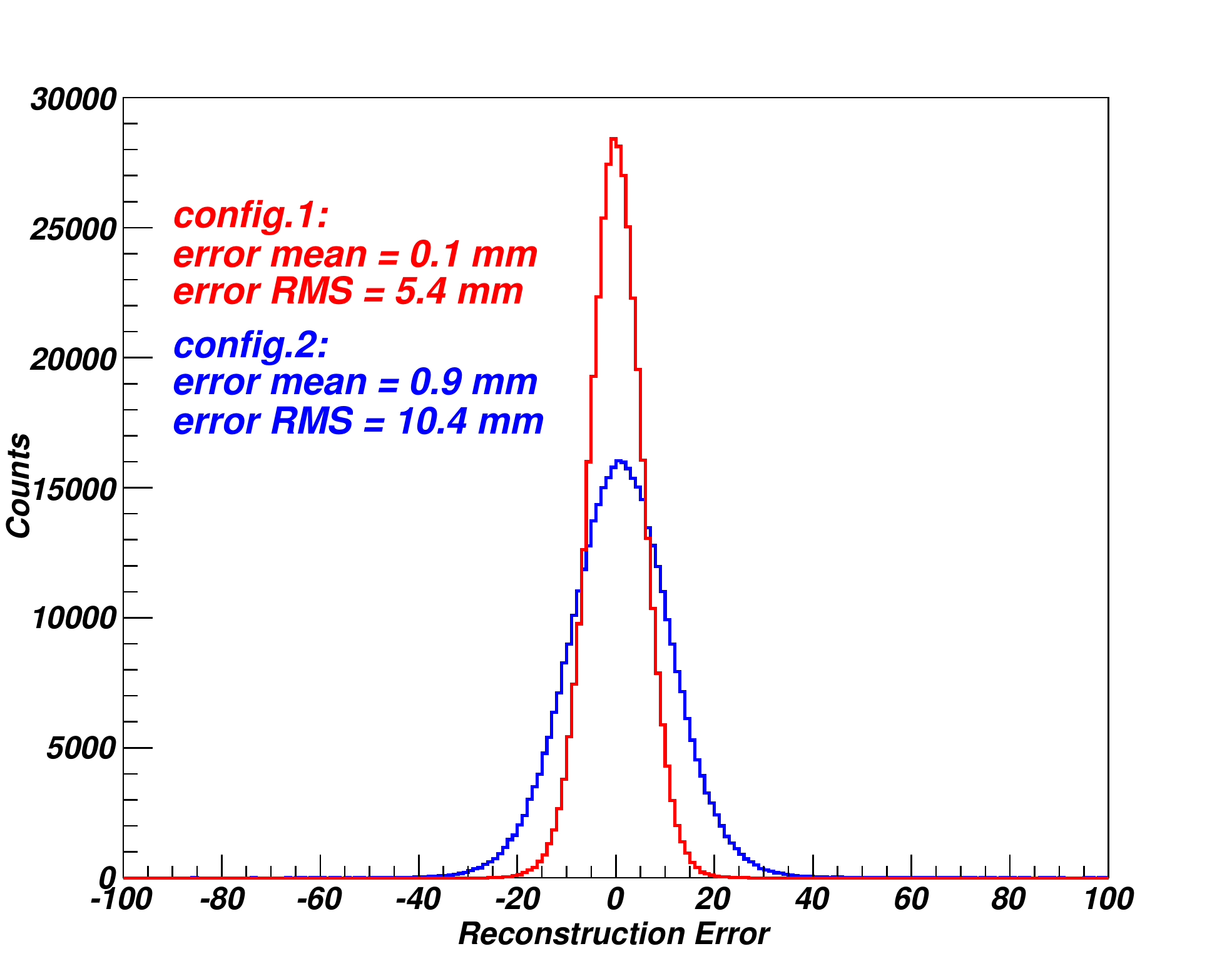}}
\caption{Reconstruction error and their spread in config.1 and config.2.}
\label{figure 8}
\end{figure}

To examine the impact of different position resolutions on the total background in these two configurations,
we define the background enhancement factor ($\triangle_{BKG}$) as:
\begin{equation}
\triangle_{BKG} = \frac{N_{reconst} - N_{ori}}{N_{ori}}
\end{equation}
$N_{ori}$ and $N_{reconst}$  stand for the number of events before and after the reconstruction in the fiducial volume respectively.
To understand the background enhancement from the PMTs with different configurations,
the background distribution as a function of radius for single-scatter electron recoil events in the energy range below 50 $\rm{keV_{ee}}$ and with $z$ between -125~mm and -925~mm,
are obtained from config.1 and config.2 respectively.
These distributions are fed into the Geant4 code to simulate the S2 signal for position reconstruction.
The event distributions before and after the position reconstruction are shown in figure 9.
The background enhancement factor after the radius selection ($r$ < 450 mm) is 2.4\% for config.1 and 3.1\% for config.2,
both are negligible compared to the original background without the position smearing.
For a uniformly distributed background, such as from the $pp$ solar neutrinos,
the background enhancement factors in the fiducial volume are about 0.4\% and 0.5\% for config.1 and config.2 respectively, which are also negligible.

\begin{figure}[htd]
\centerline{\includegraphics[width=.5\textwidth]{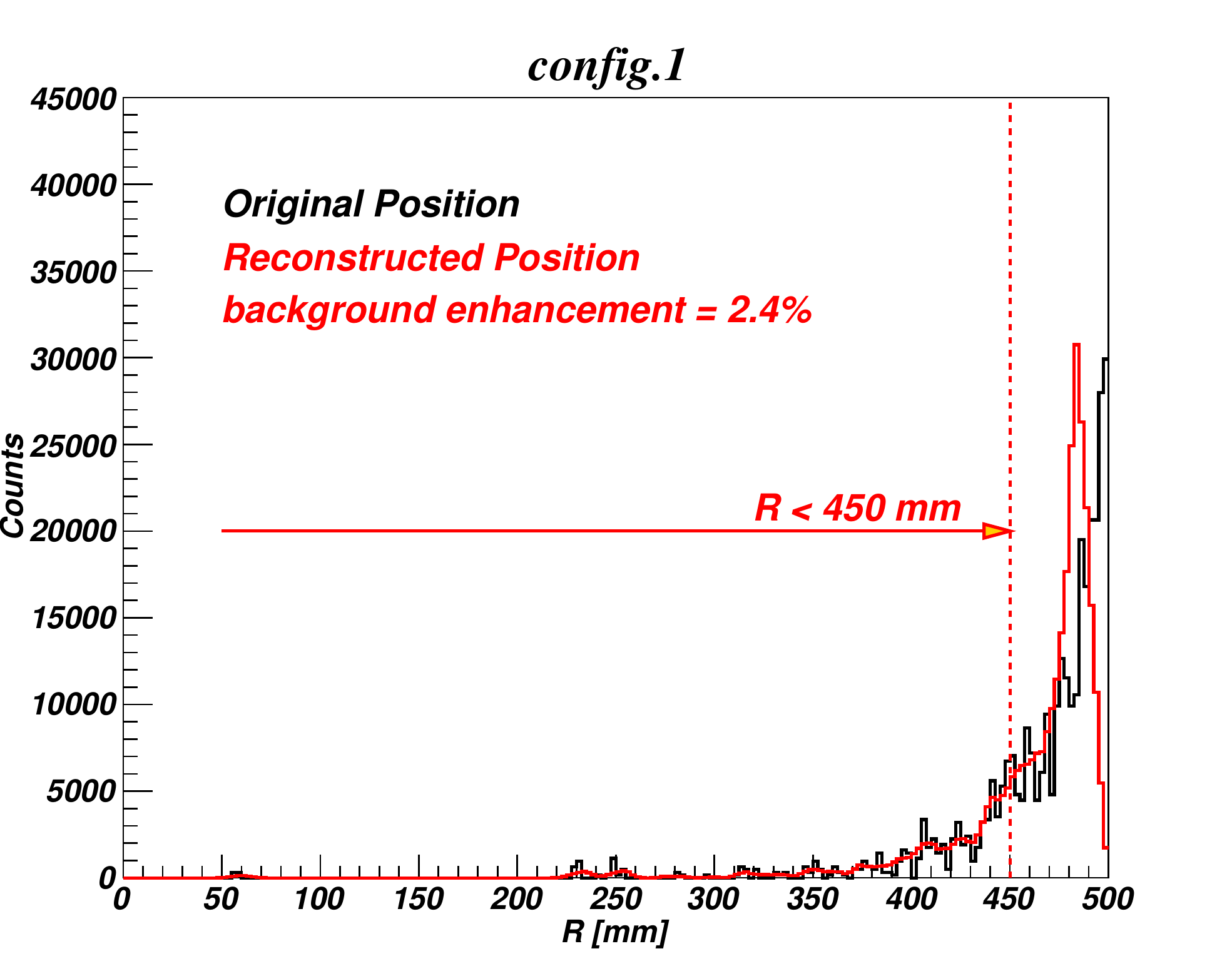} \hspace{1cm}
            \includegraphics[width=.5\textwidth]{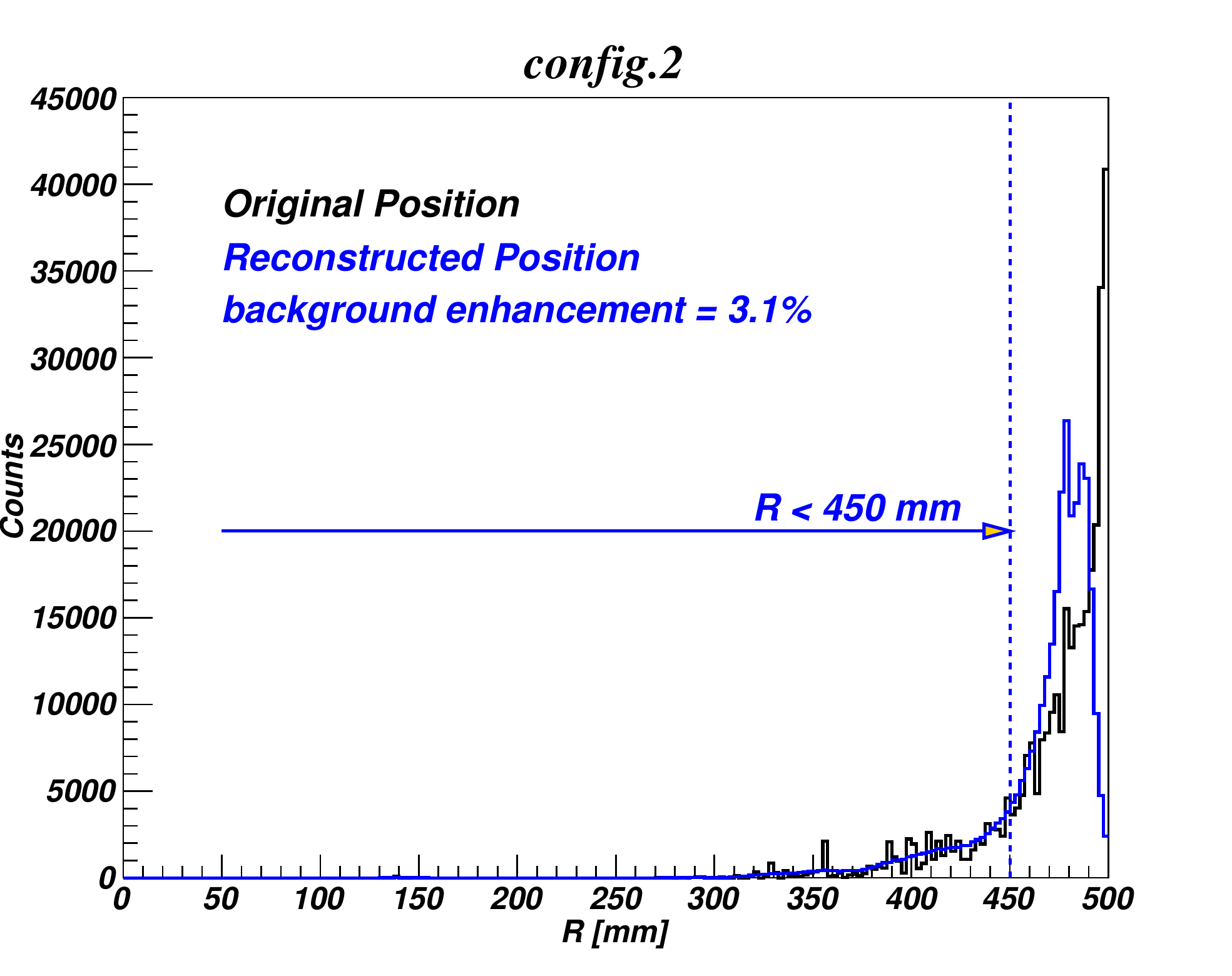}}
\caption{The original and the reconstructed background distribution from the PMTs as a function of R for both configurations.}
\label{figure 9}
\end{figure}

\section{Projected sensitivity to dark matter}

To assess the sensitivity of the LXe detector studied above, we calculated the
90\% confidence level exclusion limit for the spin-independent WIMP-nucleon elastic scattering cross section,
where WIMPs are assumed to be distributed in an isothermal halo with $v_{0}$ = 220 $\rm{km/s}$, galactic escape velocity $v_{esc}$ = 544 $\rm{km/s}$,
and a density of $\rho$ = 0.3 $\rm{GeV/cm^{3}}$ \cite{bib22}.
An average nuclear recoil acceptance of 0.35 is assumed.
The scintillation efficiency ($L_{eff}$) of nuclear recoils relative to 122~$\rm{keV}_{ee}$ gamma rays at zero field is taken from XENON100 \cite{bib23}.
Due to the large variations of light yield and background in the final detector, we consider both the best and the worst scenarios.
The worst scenario is with the lowest simulated light yield and the highest background.
The best scenario is with the highest simulated light yield and the lowest background.
The S1 resolution, governed by the Poisson fluctuation of photoelectrons (PEs) in the PMTs, is taken into account.
The energy window with S1 between 3 - 30 PEs, corresponding to the most probable energy window for WIMP scattering, is chosen for the best scenarios.
For the worst scenarios, we choose an energy window between 3 - 20 PEs for S1 to lower the expected number of background events.
The number of electron recoil background events is calculated based on the simulated values from the PMTs in table 2 and $(1.4+0.07E) \times 10^{-5}$ $\rm{events/keV_{ee}/kg/day}$
from $pp$ solar neutrinos and $^{136}$Xe double beta decays~\cite{bib14b}.
The number of nuclear recoil background events is calculated based on the simulated values from the PMTs and discussions in section 3.2. The parameters used in the sensitivity calculation are listed in table 4.

The projected sensitivity curves are shown in figure 10 for two live-years of exposure in the 1.4 ton fiducial mass based on the numbers in table 4.
The sensitivity is affected by the light yield, which depends largely on the PTFE reflectivity and the absorption length of UV photon in liquid xenon.
The choice of the different models of PMTs for the top PMT array has a small impact on the final sensitivity, especially for heavy WIMPs above 100~$\rm{GeV}$/$c^{2}$.
In the absence of a positive signal,
both detector configurations can exclude cross sections above $2(3) \times 10^{-47} \rm{cm}^{2}$ for 100~$\rm{GeV}$/$c^{2}$ WIMPs at 90\% confidence level
in the best (worst) scenarios . For light WIMPs at 10~$\rm{GeV}$/$c^{2}$, the config.1 detector can reach 0.2$(1.2)\times 10^{-45} \rm{cm}^{2}$ for the best (worst) scenarios
and the config.2 detector can reach 0.4$(2.6)\times 10^{-45} \rm{cm}^{2}$ for the best (worst) scenarios.

\begin{center}
	\begin{table}
	\caption{
    Parameters for calculating the sensitivity curves.
	The second column is the light yield while the third column is nuclear/electron recoil energy window corresponding to the light yield and S1 window.
	$N_{BKG}$ is the number of background events assuming a 99.5\% electron recoil rejection efficiency, a 35\% nuclear recoil acceptance and 1.4 ton $\times$ 2 live-years exposure.
	$\ast$  is the electron recoil events from PMTs. $\dagger$ is for the electron recoil background events from $pp$ solar neutrinos and double beta decays of ${}^{136}\rm{Xe}$. $\ddagger $ is the expected nuclear recoil background events.
	}
	\centering
	\footnotesize
	\begin{tabular}{|c|c|c|c|c|}
	\hline
		Scenario			&      Light Yield          & S1 window     & Energy window                                           &  $N_{bkg}$ \\
	             			&	$\rm{pe/keV}_{ee}$      &  [pe]         & [$\rm{keV}_{ee}$] / [$\rm{keV}_{nr}$]                   &  events/1.4~ton/2~year   \\ \hline	
		worst  of config.1  & 	1.8     	        	& 3-20			&		[1.7, 11.1] / [7.8, 37.4]				          &	 $0.10^{\ast}$ + $0.89^{\dagger}$ + $0.13^{\ddagger}$ \\
		best of config.1    & 	4.3     	        	& 3-30 			&		[0.7, 7.0] / [4.9, 25.1]				          &	 $0.02^{\ast}$  + $0.54^{\dagger}$ + $0.06^{\ddagger}$\\
		worst of config.2   & 	1.3     		        & 3-20 			&		[2.3, 15.4] / [10.1, 49.6]			              &	 $0.09^{\ast}$  + $1.35^{\dagger}$ + $0.16^{\ddagger}$\\
		best of config.2 	& 	3.2     		        & 3-30			&		[0.9, 9.4] / [5.1, 31.5]				          &	 $0.04^{\ast}$ + $0.76^{\dagger}$ + $0.08^{\ddagger}$\\
	\hline
	\end{tabular}
	\label{tab4}
	\end{table}
\end{center}

\begin{figure}[htb]
\centerline{\includegraphics[width=.8\textwidth]{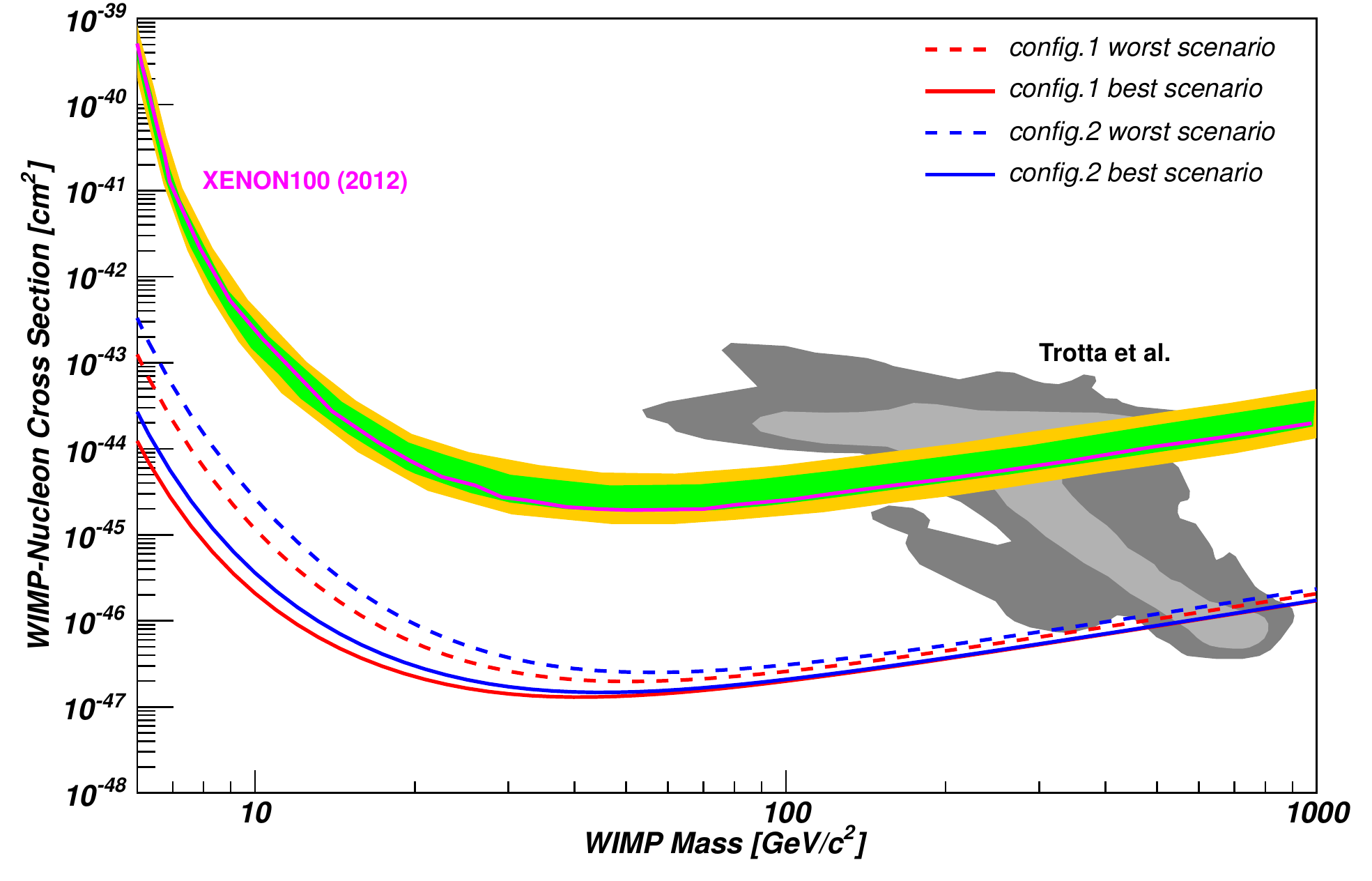}}
\caption{Projected spin-independent WIMP sensitivity for config.1 and config.2 with 90\% confidence level for a 1.4 ton $\times$ 2 live-years exposure.
For comparison, the latest result of XENON100 \cite{bib6} is also shown, together with the region preferred by the constrained minimal supersymmetric models (CMSSM)~\cite{bib24}.}
\label{figure 10}
\end{figure}

\section{Conclusion}

In this paper, we studied two configurations of time projection chambers with different top PMT arrays for a ton-scale two-phase xenon detector.
In the design, the best developed PMTs in the industry for low background dark matter experiment using liquid xenon are used.
We compared the performance of the detector with the same number of 151 Hamamatsu R8520 or R11410 PMTs for the top PMT array,
while both using 121 R11410 PMTs for the bottom PMT array. The light collection efficiency for the primary scintillation light depends heavily on the PTFE reflectivity and the absorption length of UV photons in liquid xenon.
With different input parameters, the simulated light yields in the fiducial volume at a nominal drift field of 1 kV/cm  are from 1.8 to 4.3 $\rm{pe/keV_{ee}}$
for the design with R11410 PMTs on the top and from 1.3 to 3.2 $\rm{pe/keV_{ee}}$ for the design with R8520 PMTs on the top.
The position resolutions based on the S2 hit pattern on the top PMT array are slightly different for these two configurations,
but they have a negligible effect on background enhancement of the detector.

The simulated electron recoil background rates from the PMTs for the two configurations are quite similar and they are sub-dominant compared to the background from electron scatterings of the $pp$ solar neutrinos and the double beta decays of $^{136}$Xe.
We also estimated the neutron background from the PMTs and other components to estimate the sensitivity reach of the detectors. Both designs can reach a spin-independent WIMP-nucleon elastic scattering cross section of $2(3) \times 10^{-47}$ $\rm{cm}^{2}$ for 100 $\rm{GeV}$/$c^{2}$ WIMPs
in the 1.4 ton fiducial target after two live-years of operation in the best (worst) scenarios. The key parameter affecting the sensitivity, especially for the light WIMPs,
is the light collection efficiency, which depends on the PTFE reflectivity and the UV photon absorption length in liquid xenon.

\acknowledgments

This work is supported within the PandaX experiment by the National Science Foundation of China (grant numbers: 11055003 and 11175117),
the Science and Technology Commission of Shanghai Municipality (grant number: 11PJ1405300),
and the Ministry of Science and Technology of China (grant number: 2010CB833005).
We would like to thank other members of the PandaX collaboration for valuable discussions during the work.


\begin{thebibliography}{9}

\bibitem{bib1}
G. Steigman and M. S. Turner,
\emph{Cosmological constraints on the properties of weakly interacting massive particles},
\href{http://dx.doi.org/10.1016/0550-3213(85)90537-1}
{\emph{Nucl.\ Phys.\ B.} {\bf 253} (1985) 375}

\bibitem{bib2}
G. Jungman, M. Kamionkowski and K. Griest,
\emph{Supersymmetric dark matter},
\href{http://dx.doi.org/10.1016/0370-1573(95)00058-5}
{\emph{Phys.\ Rept.} {\bf 267} (1996) 195}

\bibitem{bib3}
C. Savage et al.,
\emph{Compatibility of DAMA/LIBRA dark matter detection with other searches},
\href{http://dx.doi.org/10.1088/1475-7516/2009/04/010}
{\emph{JCAP.} {\bf 0904} (2009) 010}
\href{http://arxiv.org/abs/0808.3607v3}
{\emph\ [arXiv:0808.3607]}

\bibitem{bib4}
C. E. Aalseth et al. (CoGeNT),
\emph{Results from a search for light-mass dark matter with a P-type point contact Germanium detector},
\href{http://dx.doi.org/10.1103/PhysRevLett.106.131301}
{\emph{Phys.\ Rev.\ Lett.} {\bf 106} (2011) 131301}

\bibitem{bib5}
G. Angloher et al. (CRESST-II),
\emph{Results from 730 Kg days of the CRESST-II Dark Matter search},
\href{http://dx.doi.org/10.1140/epjc/s10052-012-1971-8}
{\emph{Eur.\ Phys.\ J.} {\bf C72} (2012) 1971}
\href{http://arxiv.org/abs/1109.0702}
{\emph\ [arXiv:1109.0702]}

\bibitem{bib6}
E. Aprile et al. (XENON100),
\emph{Dark Matter Results from 225 Live Days of XENON100 Data},
\href{http://dx.doi.org/10.1103/PhysRevLett.109.181301}
{\emph{Phys.\ Rev.\ Lett.} {\bf 109} (2012) 181301}
\href{http://arxiv.org/abs/1207.5988}
{\emph\ [arXiv:1207.5988]}

\bibitem{bib7}
Z. Ahmed et al. (CDMS),
\emph{Results from a Low-Energy Analysis of the CDMS II Germanium Data},
\href{http://dx.doi.org/10.1103/PhysRevLett.106.131302}
{\emph{Phys.\ Rev.\ Lett.} {\bf 106} (2011) 131302}
\href{http://arxiv.org/abs/1011.2482}
{\emph\ [arXiv:1011.2482]}

\bibitem{bib8}
E. Aprile et al. (XENON100),
\emph{The XENON1T dark matter search experiment},
\href{http://arxiv.org/abs/1206.6288}
{\emph\ {arXiv:1206.6288}}

\bibitem{bib9}
H. Gong, K. L. Giboni, et al.,
\emph{The Cryogenic System for the Panda-X Dark Matter Search Experiment},
\href{http://arxiv.org/abs/1207.5100}
{\emph{JINST} {\bf 8} (2013) P01002}
{\emph\ {[arXiv:1207.5100]}}


\bibitem{bib13}
D. S. Akerib et al. (LUX),
\emph{An ultra-low background PMT for liquid xenon detectors},
\href{http://arxiv.org/abs/1205.2272}
{\emph{Nucl. Instru. and Meth. A} {\bf 703} (2013) 1-6}
{\emph{[arXiv:1205.2272]}}

\bibitem{bib19}
K. Lung et al.,
\emph{Characterization of the Hamamatsu R11410-10 3-in. photomultiplier tube for liquid xenon dark matter direct detection experiments},
\href{http://dx.doi.org/10.1016/j.nima.2012.08.052}
{\emph{Nucl.\ Instrum.\ Meth. \ A} {\bf 696} (2012) 32}

\bibitem{bib10}
J. Allison et al. (GEANT4),
\emph{Geant4 developments and applications},
\href{http://dx.doi.org/10.1109/TNS.2006.869826}
{\emph{IEEE Trans.\ Nucl.\ Sci.} {\bf 53} (2006) 270}

\bibitem{bib11}
E. Aprile, K. L. Giboni, et al.,
\emph{Proportional light in a dual-phase xenon chamber},
\href{http://dx.doi.org/10.1109/TNS.2004.832690}
{\emph{IEEE Trans.\ Nucl.\ Sci.} {\bf 51} (2004) 1986}

\bibitem{bib12}
E. Aprile et al. (XENON100),
\emph{The XENON100 dark matter experiment},
\href{http://dx.doi.org/10.1016/j.astropartphys.2012.01.003}
{\emph{Astropart\ Phys.} {\bf 35} (2012) 573}

\bibitem{bib14}
Joseph A. Formaggio and C. J. Martoff,	
\emph{Backgrounds to sensitive experiments underground},
\href{http://dx.doi.org/10.1146/annurev.nucl.54.070103.181248}
{\emph{Annu.\ Rev.\ Nucl. \ Part. \ Sci.} {\bf 54} (2004) 361}

\bibitem{bib14c}
E. Aprile et al. (XENON100),
\emph{Study of the electromagnetic background in the XENON100 experiment},
\href{http://link.aps.org/doi/10.1103/PhysRevD.83.082001}
{\emph{Phys.\ Rev.\ D.} {\bf 83} (2011) 082001}

\bibitem{Titanium}
D.S. Akerib, et al.,
\emph{Radio-assay of Titanium samples for the LUX Experiment},
\href{http://arxiv.org/abs/1112.1376}
{\emph[arXiv:1112.1376v3]}

\bibitem{bib16}
E. Aprile et al. (XENON100),
\emph{Material screening and selection for XENON100},
\href{http://dx.doi.org/10.1016/j.astropartphys.2011.06.001}
{\emph{Astropart\ Phys.} {\bf 35} (2011) 43}
\href{http://arxiv.org/abs/1103.5831}
{\emph[arXiv:1103.5831]}

\bibitem{bib14a}
K. Abe et al. (XMASS),	
\emph{Distillation of liquid xenon to remove krypton},
\href{http://dx.doi.org/10.1016/j.astropartphys.2009.02.006}
{\emph{Astropart\ Phys.} {\bf 31} (2009) 290}

\bibitem{bib14b}
K. Arisaka et al.	
\emph{Expected sensitivity to galactic/solar axions and bosonic Super-WIMPs based on the Axio-electric effect in liquid xenon dark matter detectors},
{\emph{Astropart\ Phys.} {\bf 44} (2013) 59-67}
\href{http://arxiv.org/abs/1209.3810}
{\emph\ arXiv:1209.3810 [astro-ph.CO]}

\bibitem{SOURCE}
M. Carson et al,
\emph{Neutron background in large-scale xenon detectors for dark matter searches},
\href{http://dx.doi.org/10.1016/j.astropartphys.2004.05.001}
{\emph{Astropart\ Phys.} {\bf 21} (2004) 667}

\bibitem{Claudio F. P. Silva}
C.F. P. Silva,
\emph{PTFE reflectance measurements, modeling and simulation for xenon detectors,}
\href{https://indico.cern.ch/getFile.py/access?contribId=169&sessionId=21&resId=0&materialId=slides&confId=102998}
{\emph\ University of Coimbra LIP Laboratory, presentation at TIPP 2011, 11 June 2011}

\bibitem{Baldini}
A. Baldini et al,
\emph{Absorption of scintillation light in 100 L liquid xenon $\gamma$ ray detector and expected detecotr performance},
{\emph{Nucl\ Instr\ and \ Meth A} {\bf 545} (2005) 753}

\bibitem{bib17}
E. Aprile et al. (XENON10),
\emph{Design and performance of the XENON10 dark matter experiment},
\href{http://dx.doi.org/10.1016/j.astropartphys.2011.01.006}
{\emph{Astropart\ Phys.} {\bf 34} (2011) 679}
\href{http://arxiv.org/abs/1001.2834}
{\emph\ [arXiv:1001.2834]}

\bibitem{bib18}
K. Ni,
\emph{Development of a liquid xenon time projection chamber for the XENON dark matter Search },
{\emph{Ph.D. Thesis, Columbia University} (2006)}
\href{http://jinst.sissa.it/jinst/theses/2006_JINST_TH_007.jsp}
{\emph {JINST} {TH} {2006} {007}}


\bibitem{hamamatsu}
Hamamatsu Inc., Private communications.

\bibitem{arxiv:1207.5432}
E. Aprile et al.,
\emph{Measurement of the Quantum Efficiency of Hamamatsu R8520 Photomultipliers at Liquid Xenon Temperature},
{\emph{JINST} {\bf 7} (2012) P10005}
\href{http://arxiv.org/abs/1207.5432}
{\emph\ [arXiv:1207.5432]}

\bibitem{bib20}
M. Szydagis et al.,
\emph{NEST: a comprehensive model for scintillation yield in liquid xenon},
\jinst {6} {2011} {P10002}

\bibitem{ZEPLIN-III}
V.N. Solovov et al.,
\emph{Position reconstruction in a dual phase xenon scintillation detector},
\href{http://arxiv.org/pdf/1112.1481v2}
{\emph\ [arXiv:1112.1481v2]}

\bibitem{bib21}
E. Aprile et al.,
\emph{Simultaneous measurement of ionization and scintillation from nuclear recoils in liquid xenon for a dark matter experiment},
\href{http://dx.doi.org/10.1103/PhysRevLett.97.081302}
{\emph{Phys.\ Rev.\ Lett.} {\bf 97} (2006) 081302}

\bibitem{bib22}
M. G. Abadi1, J. F. Navarro and M. Steinmetz,
\emph{Stars beyond galaxies: the origin of extended luminous haloes around galaxies},
\href{http://dx.doi.org/10.1111/j.1365-2966.2005.09789.x}
{\emph{Mon.\ Not.\ R.\ Astron. \ Soc.} {\bf 375} (2007) 755}

\bibitem{bib23}
E. Aprile et al. (XENON100),
\emph{Dark matter results from 100 live days of XENON100 data},
\href{http://dx.doi.org/10.1103/PhysRevLett.107.131302}
{\emph{Phys.\ Rev.\ Lett.} {\bf 107} (2011) 131302}

\bibitem{bib24}
R. Trotta1, F. Feroz, M. Hobson, L. Roszkowski and R. R. Austri,
\emph{The impact of priors and observables on parameter inferences in the constrained MSSM},
\href{http://dx.doi.org/10.1088/1126-6708/2008/12/024}
{\emph{J.\ High Energy Phys.} {\bf 12} (2008) 024}

\end{thebibliography}
\end{document}